\documentclass[notitlepage,superscriptaddress,aps,prb,twocolumn,noeprint,longbibliography]{revtex4-2}
\usepackage{amsmath}
\usepackage{amssymb}
\usepackage[sort&compress]{natbib}
\usepackage{hyperref}
\setcitestyle{numbers,comma,square}
\usepackage{upgreek}
\usepackage{mathtools}
\usepackage{graphicx}
\usepackage{afterpage}
\usepackage{amssymb}
\usepackage{epstopdf}
\usepackage{xr}
\usepackage{tikz}
\usepackage{color}
\usepackage{bm}
\usepackage[caption=false,position=top,singlelinecheck=off,justification=raggedright]{subfig}

\graphicspath{{Images/}}

\DeclarePairedDelimiter\bra{\langle}{\rvert}  
\DeclarePairedDelimiter\ket{\lvert}{\rangle}       
\DeclarePairedDelimiterX\braket[2]{\langle}{\rangle}{#1 \delimsize\vert #2}

\begin{document}

\title{Trajectory-Resolved Weiss Fields for Quantum Spin Dynamics}
\author{S. E. Begg}
\affiliation{King's College London, Strand, London WC2R 2LS, United Kingdom}
\affiliation{Asia Pacific Center for Theoretical Physics, Pohang 37673, Korea}
\author{A. G.  Green}
\affiliation{London Centre for Nanotechnology, University College London,
Gordon St., London, WC1H 0AH, United Kingdom}
\author{M. J. Bhaseen}
\affiliation{King's College London, Strand, London WC2R 2LS, United Kingdom}
\date{\today}
\begin{abstract}
We explore the dynamics of quantum spin systems in two and three
dimensions using an exact mapping to classical stochastic processes. In
recent work we explored the effectiveness of sampling
around the mean field evolution as determined by a stochastically
averaged Weiss field. Here, we show that this approach can be
significantly extended by sampling around the instantaneous Weiss
field associated with each stochastic trajectory taken separately. This
trajectory-resolved approach incorporates sample to sample fluctuations
and allows for longer simulation times. We demonstrate the utility of
this approach for quenches in the two-dimensional and three-dimensional
quantum Ising model. We show that the method is particularly
advantageous in situations where the average Weiss-field vanishes, but
the trajectory-resolved Weiss fields are non-zero. We discuss the
connection to the gauge-P phase space approach, where the
trajectory-resolved Weiss field can be interpreted as a gauge degree of freedom.
\end{abstract}
\maketitle

\section{Introduction}
Quantum spin systems play a prominent role in the field of many-body
physics with diverse applications ranging from quantum magnetism to
non-equilibrium dynamics \cite{Polkovnikov2011,Heyl2017,Abanin2019}.
They also play a crucial role in the development
of theoretical techniques ranging from methods of integrability \cite{Rigol2007,Rigol2008,Essler2012,Cazalilla2011,Batchelor2016} to
numerical algorithms \cite{White1992,White1993,White2004,Vidal2004,Haegeman,Carleo2017}. 
A notable challenge is the theoretical
description of two and three-dimensional quantum spin systems, where
the lack of integrability and the dimension of the Hilbert space stymies progress. The
growth of quantum entanglement in real-time dynamics also impedes the
description of non-equilibrium phenomena beyond short timescales; this is particularly severe in higher dimensions, as discussed in Refs. \cite{Czarnik2019,Paeckel2019,Hubig2019,Hubig2020,Schmitt2020}. In addition, tensor network representations are computationally less tractable than in one-dimension, with the number of network contractions scaling exponentially with the system size \cite{Schuch2007}; this significantly reduces the accessible time-scales. Recent progress has been made using machine learning techniques \cite{Carleo2017,Schmitt2020,Burau2021}, and via semi-classical approaches based on the truncated Wigner approximation \cite{Schachenmayer2015,Schachenmayer2015b,Wurtz2018,Khasseh2020}. 

Recently, a stochastic approach to quantum spin systems has been
developed, based on a Hubbard--Stratonovich decoupling of the exchange
interactions \cite{Hogan2004,Galitski2011,Ringel2013,DeNicola2019,DeNicola2019long,Begg2019,DeNicola2019euclid,Begg2020,DeNicola2021}. This approach provides an exact reformulation of the quantum dynamics in terms of classical stochastic processes. Quantum
expectation values are computed by averaging over independent
stochastic trajectories, and the method can be applied in arbitrary
dimensions. In recent work we showed that one could extend the
accessible simulation times in this approach through the use of an
effective Weiss-field \cite{Begg2020}. Similar conclusions have also been inferred using saddle point techniques \cite{DeNicola2019euclid,DeNicola2021}. These approaches allow one to expand around the
stochastically averaged time-evolution in order to reduce the need
for large stochastic fluctuations. We further discussed \cite{Begg2020} the connection to the gauge-P phase space formulation \cite{Deuar2002}, and the possibility to make efficient choices of gauge. This complements a large body of phase space techniques which have emerged in recent years   \cite{Drummond1980,Deuar2002,Polkovnikov2003b,Polkovnikov2003a,Barry2008,Ng2011,Ng2013,Mandt2015,Wuster2017,Deuar2021,Steel1998,Polkovnikov2010,Schachenmayer2015,Schachenmayer2015b,Wurtz2018,Khasseh2020,Huber2020,Verstraelen2018,Verstraelen2020,Hush2009,Ng2022,Kiesewetter2022}.  

In this work, we show that the stochastic approach can be
significantly enhanced by allowing the Weiss field to be determined on a
trajectory by trajectory basis. This allows one to generalize the expansion around a single mean-field trajectory and incorporate sample to sample fluctuations. This trajectory-resolved approach is particularly
advantageous in situations where the stochastically averaged Weiss
field vanishes. We show that the approach can lead to longer
simulation times for a range of quantum quenches in both two and three dimensions. In particular, we demonstrate an exponential improvement in the sampling efficiency. This establishes the technique as a viable tool for the simulation of non-equilibrium quantum spin systems beyond one-dimension. 

The layout of this paper is as follows. In Section  \ref{sec:stochasticformalism} we provide an overview of the stochastic approach. In Sections \ref{sec:averageweiss} and \ref{sec:trajectoryweiss} we discuss the use of trajectory-averaged and trajectory-resolved Weiss fields respectively. In Section  \ref{sec:simulations} we use these Weiss fields to simulate quantum quenches in the two and three-dimensional quantum Ising model. In Sections \ref{sec:stochasticnorm} and \ref{sec:growthfluct} we discuss the growth of stochastic fluctuations under time-evolution. We conclude in Section  \ref{sec:conclusion} and provide an appendix on the associated links \cite{Begg2020} to the gauge-P phase space formalism \cite{Deuar2001,Deuar2002,Drummond2003,Drummond2004,Barry2008,Ng2013,Wuster2017}. We also provide details of our numerical simulations.

\section{Stochastic Formalism} \label{sec:stochasticformalism}
In this section we recall the principal features of the stochastic approach to quantum spin systems \cite{Hogan2004,Galitski2011,Ringel2013,DeNicola2019,DeNicola2019long,DeNicola2019euclid,Begg2019,Begg2020,DeNicola2021}. The method is applicable to
generic quadratic spin Hamiltonians in arbitrary dimensions: 
\begin{equation} \hat{H} = -\frac{1}{2} \sum_{jkab}  J^{ab}_{jk} \hat{S}^a_j  \hat{S}^b_k - \sum_j h^a_j \hat{S}^a_j , \label{eq:heisenbergham}\end{equation}
where $J^{ab}_{jk}$ is the exchange interaction between sites $j$ and $k$ and
$h^a_j$ is an applied magnetic field. The spin operators, $\hat{S}^a_j$, obey the canonical commutation relations $[\hat{S}^a_j,\hat{S}^b_k] = i \epsilon^{abc}\delta_{jk}\hat{S}^c_k,$ where $a,b  \in  \{x,y,z\}$ label the spin components, $\epsilon^{abc}$ is the antisymmetric symbol and we set $\hbar = 1$.  

The dynamics of the quantum spin system is encoded in the
time-evolution operator $\hat{U}(t) = \mathbb{T}e^{-i \int_{0}^{t} \hat{H}(t) dt},$ where $\mathbb T$ denotes
time-ordering. Decoupling the interactions by means of a
Hubbard--Stratonovich transformation \cite{Hubbard1959} over auxiliary fields $\varphi_j$
one obtains 
\begin{align}  \hat{U}(t_f,t_i)\! =\! \mathbb{T} \int \! \mathcal{D}\varphi  ~ e^{- S[\varphi] +  i\int\limits_{\mathclap{t_i}}^{\mathclap{t_f}}\! dt \sum_{ja} \Phi_j^a \hat{S}^a_{j}} ,\label{eq:HStransf} \end{align}
where $\Phi_j^a =
\frac{1}{\sqrt{i}}\varphi_j^a + h_j^a\in {\mathbb C}$ and the integration is performed over all paths with $\mathcal{D}\varphi=\prod_{ja} {\mathcal D}\varphi^a_j$. The parameter $\Phi_j^a$ plays the role
of an effective, complex magnetic field. 
 In writing Eq. (\ref{eq:HStransf}) we define the so-called
{\em noise action}  
\cite{DeNicola2019long,DeNicola2019euclid,Begg2020,DeNicola2021}, 
\begin{align} S[\varphi] = \frac{1}{2}
  \int_{t_i}^{t_f} dt \sum_{jkab} \varphi^a_j~(J^{-1})^{ab}_{jk}
  \varphi^b_{k} \label{eq:whitenoisemeasure},\end{align}
 which allows one to regard the fields $\varphi$ as correlated random noises with the Gaussian measure $\mathcal{D}\varphi  ~ e^{- S[\varphi]}$. 
The time-evolution operator (\ref{eq:HStransf}) can be recast as \begin{align}\hat{U}(t) = \big\langle  \mathbb{T} e^{-i \int_0^t \hat{H}^s(t') dt' } \big\rangle_{\varphi} \label{eq:evo2}\end{align}  where  $\hat{H}^s \equiv - \sum_{ja}
\Phi_j^a \hat{S}^a_j$ is referred to as the stochastic Hamiltonian  
and $\langle ... \rangle_{\varphi}$ denotes averaging over the
Gaussian noise variables. Eq. (\ref{eq:evo2}) motivates the introduction of the stochastic evolution operator $\hat{U}^s(t) =\mathbb{T} e^{-i \int_0^t \hat{H}^s(t') dt' }$ and the stochastic state $\ket{\psi^s(t)} = \hat{U}^s(t)\ket{\psi^s(0)}$. 
As can be seen from Eq. (\ref{eq:whitenoisemeasure}), the exchange interactions enter the representation via the correlations of the
noise. For both analytical and numerical calculations it is
convenient to convert these noises to 
Gaussian white noise by
diagonalizing the noise action 
(\ref{eq:whitenoisemeasure}) \cite{Ringel2013,DeNicola2019,DeNicola2019long,Begg2019,DeNicola2019euclid,Begg2020,DeNicola2021}. 
 Specifically, we decompose the original fields $\varphi_j^a$ in terms of white noise variables, $\phi^a_k$, as  $\varphi_j^a =  \sum_{kb} O^{ab}_{jk} \phi^b_k,$ where $\bm{O}^T\bm{J}^{-1}\bm{O} = \bm{1} $ and the bold symbols indicate matrices. 
The time-evolution operator (\ref{eq:evo2}) can be further simplified by means of a so-called disentangling transformation \cite{Ringel2013}. This eliminates the time-ordering operation by introducing a new set of variables $\xi^a_j$: \begin{align} & \hat{U}^{s}_j(t) = \mathbb{T} e^{-i \int_0^t \hat{H}^s_j dt}  =    e^{\xi^+_j(t) 
  \hat{S}^+_j} e^{\xi^z_j(t) \hat{S}^z_j} e^{\xi^-_j(t) \hat{S}^-_j}. \label{eq:disentangle}\end{align}
The variables $\xi$ satisfy the stochastic differential equations (SDEs) \cite{Ringel2013,DeNicola2019,DeNicola2019long,DeNicola2019euclid,Begg2019,Begg2020}
\begin{subequations}
\label{eq:SDEs}
\begin{align}
 & -i \dot{\xi}^+_j = \Phi^+_j + \Phi^z_j \xi^+_j - \Phi^-_j \xi^{+^2}_j,        \label{eq:plus} 
\\ & -i \dot{\xi}^z_j = \Phi^z_j-2 \Phi^-_j \xi^+_j, \label{eq:zequat} \\ &                                                       
-i \dot{\xi}^-_j = \Phi^-_j e^{\xi^z_j}.  \label{eq:mininit}  
\end{align}  
\end{subequations}
where $\Phi^{\pm}_j = \frac{1}{2} (\Phi^x_j \mp i \Phi^y_j)$ and $\xi_j^a(0) = 0$. 
Time-evolution can therefore be achieved by solving these
SDEs numerically. In order to evaluate generic local observables
$\langle \hat{{\mathcal O}(t)} \rangle =
\bra{\psi(0)} \hat{U}^{\dagger} \hat{{\mathcal O}} \hat{U} \ket{\psi(0)}$, 
one may decouple the forwards and backwards time-evolution
operators independently \cite{DeNicola2019}. Expectation values therefore reduce to averages over these classical stochastic variables:  
\begin{align}\langle \hat{{\mathcal O}}(t) \rangle =
\bra{\psi(0)} \hat{U}^{s \dagger}_{\tilde{\varphi}}(t) \hat{{\mathcal O}} \hat{U}^s_{\varphi}(t) \ket{\psi(0)}_{\varphi,\tilde{\varphi}},\end{align} 
where $\varphi$ and $\tilde\varphi$ correspond to the forwards and backwards evolution
respectively as indicated by the subscripts on the evolution operators. 
Employing the disentangling transformation given in (\ref{eq:disentangle}) one may recast this in the form $\langle \hat{{\mathcal O}}(t) \rangle = \langle f(\xi, \tilde{\xi}) \rangle_{\varphi, \tilde{\varphi}},$  where
the function $f(\xi, \tilde{\xi})$ depends on the operator $\hat{{\mathcal O}}$. Evaluating this expression involves solving the SDEs (\ref{eq:SDEs}) and computing the classical average over realizations of
the stochastic process. In general, the variables $\xi$ grow without bound as they approach coordinate singularities, leading to a failure of numerical integration schemes. This can be seen by considering the action of $\hat{U}^s(t)$ on an initial down-state $\ket{\downarrow}:$
\begin{align}
\hat{U}^s(t)\ket{\downarrow}  = e^{-\xi^z_j(t)/2}(\ket{\downarrow} + \xi_j^+(t) \ket{\uparrow}) \label{eq:blochsphere}.
\end{align}
It is evident that the up-state $\ket{\uparrow}$ is associated with the diverging quantity $|\xi_j^+(t)| \rightarrow \infty$.
As discussed in Ref. \cite{Begg2019}, these singularities can be eliminated by
a suitable parameterization of the Bloch sphere.  Specifically, one introduces a second coordinate patch $\tilde{\xi}$ for the Bloch sphere, which instead has the coordinate singularity at the down-state $\ket{\downarrow}$. Singularities can therefore be avoided by mapping between the coordinate patches whenever the evolution crosses the equator of the Bloch sphere, associated with $|\xi^+_j| = 1$. Note that by using the parametrization (\ref{eq:blochsphere}), the variable $\xi^-$ in (\ref{eq:mininit}) is not required. 
 Since any spin state can be obtained as a rotation from a down-state $\ket{\downarrow}$, the parametrization (\ref{eq:blochsphere}) can be used generically, provided one introduces a state preparation protocol. For example, one may rotate the down-state $\ket{\downarrow}$ to the initial state  $\ket{\psi_j(0)} = \hat{U}^s_j(0,-\delta)\ket{\downarrow},$ over some arbitrary time-interval $\delta$ before $t = 0$. In practice, this evolution need not be computed since we are only interested in calculating the time-evolution from $t=0$; the initial conditions on $\xi$ are constrained by the initial state according to $\hat{U}^s_j(\xi_j(0)) \ket{\downarrow}= \ket{\psi_j(0)}.$ Entangled initial states can be treated by introducing a probability distribution over these initial conditions \cite{Begg2019}. Further extensions also exist for combining the SDEs (\ref{eq:SDEs}) with matrix product states \cite{Begg2019}. Although this results in improvements in 1D, we do not pursue this here; for 2D and 3D systems the number of tensor network contractions scales exponentially with the system size \cite{Schuch2007} and no advantage is gained. Instead, we turn our attention to the use of Weiss fields.

\begin{figure}[t]
\includegraphics[width = 9cm]{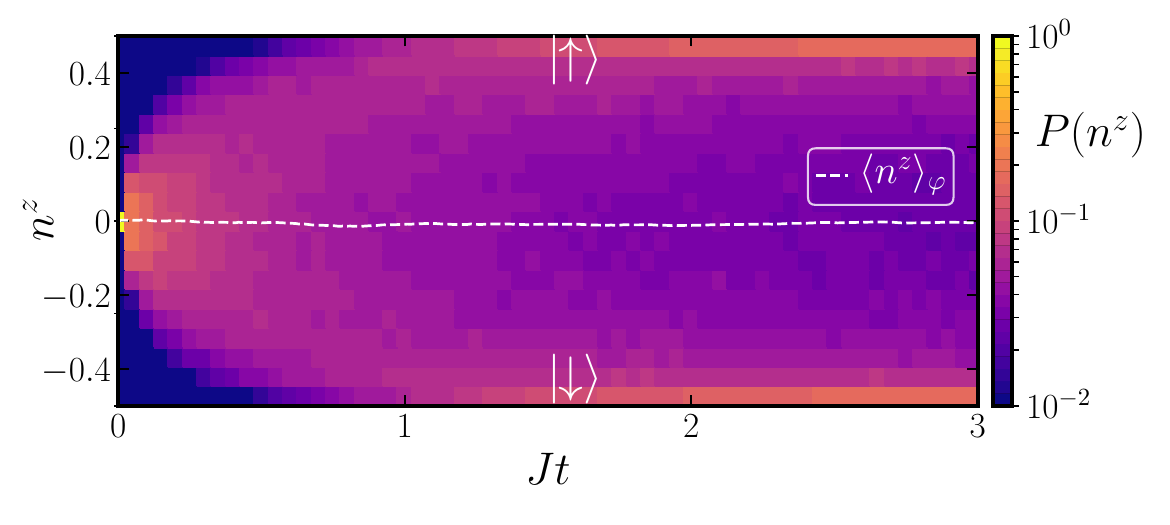}
\caption{Time-evolution of the distribution of stochastic trajectories for the z-component of spin
$P(n^z)$. We consider a quantum quench in the two-dimensional quantum Ising
model from the disordered to the ordered phase. We examine a $3\times
3$ array of spins which are initialized in the state $\ket{+}$ which is
fully-polarized along the positive $x$-direction, and time-evolve with
$\Gamma/J=0.1$. The spins evolve rapidly over the Bloch sphere and
gather at the poles, as highlighted by the bright regions. It can be
seen that the trajectory-averaged position $\langle n^z \rangle_{\varphi}$ (dashed line) doesn't
approximate the trajectory dynamics. This motivates the use of
trajectory-resolved Weiss fields in the stochastic approach. The data
correspond to all of the spins with ${\mathcal N}=1000$ stochastic
samples.}
\label{fig:distn}
\end{figure} 
\section{Trajectory-Averaged Weiss Field} \label{sec:averageweiss}
In recent work in both Euclidean \cite{DeNicola2019euclid} and real-time evolution \cite{Begg2020,DeNicola2021} it has
been noted that the sampling efficiency can be improved by shifting the
Hubbard–Stratonovich fields by a constant at each time slice t: $\varphi(t) \rightarrow \varphi(t) + \Delta \varphi(t).$
As discussed in Ref. \cite{Begg2020} it is convenient to parameterize this as $\Delta \varphi^a_j(t) = \sqrt{i} \sum_{kb} J^{ab}_{jk} m_k^b(t).$ This introduces a term of the form $\sum_{kb} J^{ab}_{jk} m_k^b(t)\hat{S}^a_{j}$ into the stochastic Hamiltonian, which can be interpreted as a Weiss field:  
\begin{align} \hat{H}^{s}(t) = &- \sum_{ja}  h_j^a \hat{S}^a_{j} - \sum_{jkab} J^{ab}_{jk} m_k^b \hat{S}^a_{j} \label{eq:hstochwithgauge} \\ & - \frac{1}{\sqrt{i}} \sum_{ja}  \varphi_j^a   \big(\hat{S}^a_j - m_j^a \hat{\mathbb{I}} \big) + \frac{1}{2} \sum_{jkab}  J_{jk}^{ab}m^{a}_{j}m^{b}_{k} \hat{\mathbb{I}} , \nonumber \end{align}
where $\hat{\mathbb{I}}$ is the identity operator.
In writing (\ref{eq:hstochwithgauge}) we have included $\hat{\mathbb{I}}$-dependent terms in the definition of the stochastic Hamiltonian, to ensure that the Gaussian measure is still in the original form (\ref{eq:whitenoisemeasure}). 
The Weiss field allows one to sample the noise fluctuations around a single deterministic ‘mean-field’ trajectory determined by $m^a_j(t)$. This can be selected to reduce stochastic
fluctuations. In previous work \cite{Begg2020}, we set this equal to the trajectory-averaged Weiss field $m_j^a = \langle n_j^a \rangle_{\varphi}$, where  \begin{align} n_j^a =   \frac{\bra{\psi^s(t)}\hat{S}_j^a\ket{\psi^s(t)}}{|\psi^s(t)|^2} \label{eq:nz} \end{align} is the expectation value of the $a$-component of the spin. 
 In writing (\ref{eq:nz}), we include the normalization factor $|\psi^s(t)|^2$ since the stochastic state $\ket{\psi^s(t)}$ is un-normalized. Eq. (\ref{eq:nz}) can be calculated self-consistently from a relatively small number of
trajectories \cite{Begg2020}; we use four iterations of $\mathcal{O}(10^3)$ samples. The stochastic sampling is then carried out around the
deterministic trajectory determined by the first two terms in Eq. (\ref{eq:hstochwithgauge}). The trajectories are re-weighted \textit{via} the non-Hermitian term in (\ref{eq:hstochwithgauge}). This procedure therefore corresponds to a form of importance sampling \cite{DeNicola2019euclid,DeNicolaThesis}. As discussed above, without loss of generality we may consider the SDEs starting from an initial spin-down state: 
\begin{align}
 & -i \dot{\xi}^+_j = \Phi^+_j + \Phi^z_j \xi^+_j - \Phi^-_j \xi^{+^2}_j,        \label{eq:plus2} 
\\ & -i \dot{\xi}^z_j = \Phi^z_j-2 \Phi^-_j \xi^+_j + \frac{2}{\sqrt{i}} \sum_{a}  \varphi_j^a  m_j^a +   \sum_{kab}  J_{jk}^{ab}m^{a}_{j}m^{b}_{k}, \label{eq:zequat2}  
 \end{align}                    
where the $\hat{\mathbb{I}}$-dependent terms in (\ref{eq:hstochwithgauge}) enter into the evolution of $\xi^z_j$ \cite{Begg2020}
and the effective magnetic field becomes
\begin{align} \Phi_j^a =
\frac{1}{\sqrt{i}} \varphi_j^a + h_j^a + \sum_{kb}
J_{jk}^{ab}m_k^b. \label{eq:neweff}
\end{align}
It can be seen that this consists of the applied magnetic fields $h_j^a$, the Hubbard--Stratonovich field $\varphi_j^a$, and the Weiss field contribution. 
 At this stage, we note that the $\mathcal{O}(m^2)$ terms in (\ref{eq:hstochwithgauge}) and (\ref{eq:zequat2}) may be safely ignored since they result in a deterministic phase for the stochastic state $\ket{\psi^s(t)}$. 

While the trajectory-averaged Weiss field has been shown to improve simulation times for a range of quenches \cite{Begg2020}, it decays to zero over time; the trajectories spread out over the Bloch sphere. The situation can be summarized by considering a quench of the 2D quantum Ising model with nearest neighbor interactions
\begin{equation} 
\hat{H}_I =   - \frac{J}{2} \sum_{\langle ij \rangle}   \hat{S}^z_i \hat{S} ^z_{j}  - \Gamma \sum_{j=1}^{N}  \hat{S}^x_j, \label{eq:TFIM} 
\end{equation}
 where we  set $J = 1$ and use periodic boundary conditions.  In Fig.~\ref{fig:distn} we show the distribution of $n_j^z$ over time following a quench from the disordered to the
ordered phase. Over time the trajectories spread out over the Bloch sphere, accumulating at either pole. However, the mean
behavior  $\langle n_j^z \rangle_{\varphi}$ remains zero and fails to approximate the trajectories. Large non-Hermitian fluctuations are
therefore required, since the noise must drive the state towards the poles. In Section \ref{sec:trajectoryweiss} we address
this issue by allowing the Weiss field to vary on a trajectory by
trajectory basis. In Appendices \ref{sec:trajproof} and \ref{sec:driftgauges}  we give two independent derivations of the approach.

\section{Trajectory-Resolved Weiss Field}\label{sec:trajectoryweiss}
In order to treat problems where the trajectory-averaged Weiss field
vanishes, we consider the instantaneous Weiss field for each
trajectory taken separately. We restrict our attention to the quantum Ising model but discuss the application to more general models of the form (\ref{eq:heisenbergham}) in the Appendices. Specifically, we set $m^z_j(t) = n_j^z(t)$ where $n_j^z(t)$ is the instantaneous value of the z-component of the spin
for a single trajectory. 
In this approach, each trajectory develops a different Weiss field due to the effect of the accumulated noise. This may be regarded as a form of feedback \cite{Hush2009,Ng2022,Kiesewetter2022}, in which quantum fluctuations recailibrate the reference trajectory. The trajectory-resolved Weiss fields do not necessarily decay to zero at long times. 
In Section \ref{sec:growthfluct}, we show that the use of these Weiss fields allows samples to contribute more equally to quantum expectation
values. The trajectory dependence of the Weiss field also changes the effect of $\frac{1}{2} \sum_{ij}  J_{ij} m_i  m_j \hat{\mathbb{I}}$ in (\ref{eq:hstochwithgauge}) on the stochastic state $\ket{\psi^s(t)}$; it goes from a removable deterministic phase  to a stochastic phase that determines how trajectories `interfere'.

\begin{figure}[t]
\subfloat{\includegraphics[width =8.8cm]{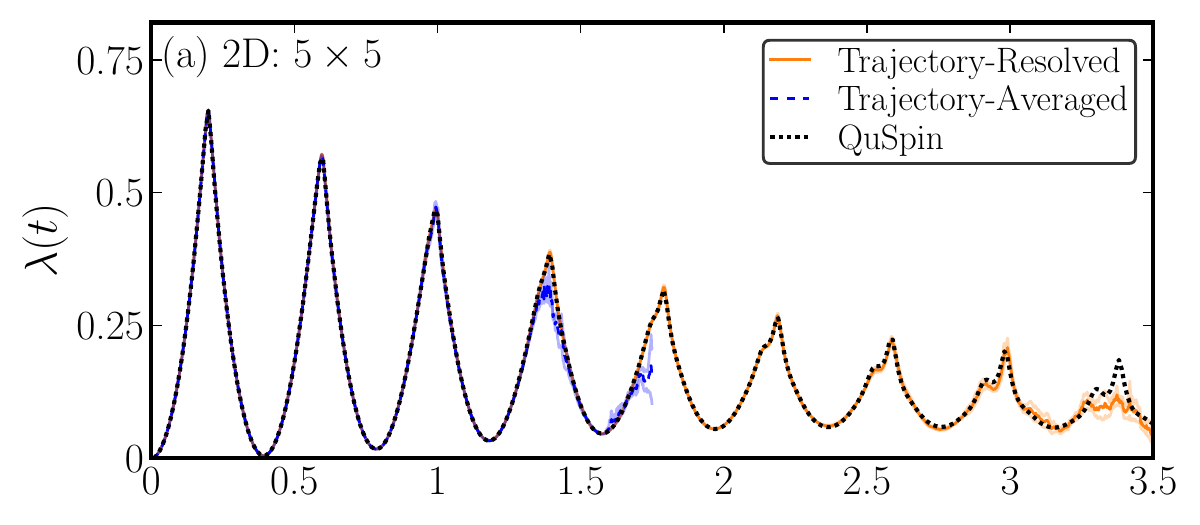}}

\vspace{-0.5cm}
\subfloat{\includegraphics[width =8.8cm]{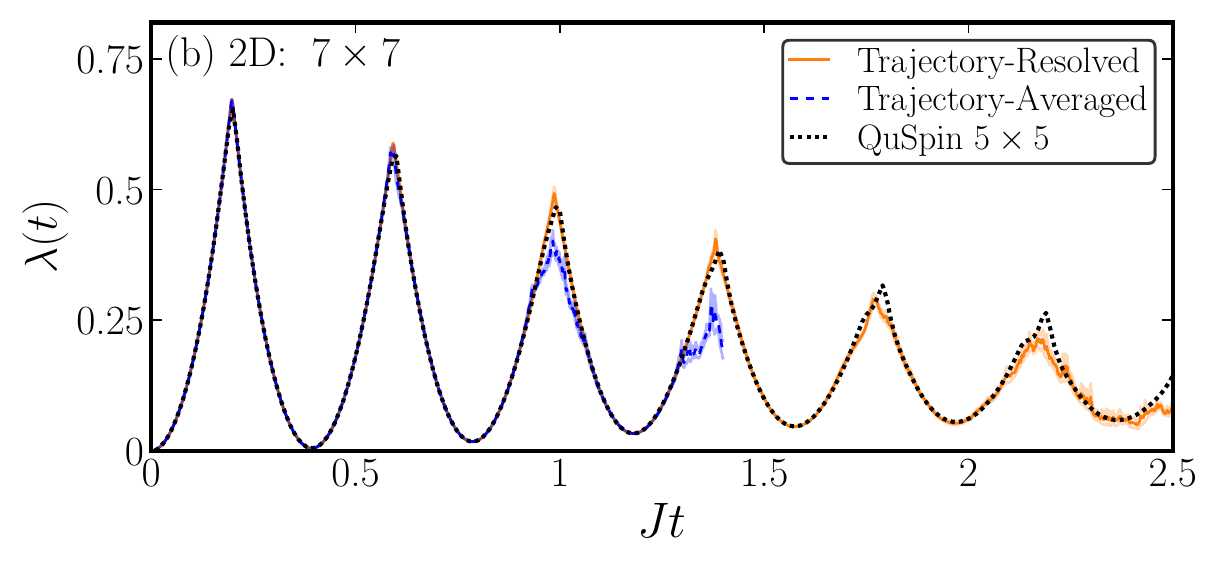}}
\caption{Time-evolution of the Loschmidt rate function $\lambda(t)$
following a quench in the two-dimensional quantum Ising model
on (a) a $5\times 5$ lattice and (b) a $7 \times 7 $ lattice. The spins
are initialized in the state $\ket{\psi(0)} = \frac{1}{\sqrt{2}}\Big( \ket{\Downarrow} + \ket{\Uparrow} \Big)$ corresponding to the superposition of degenerate ground states and time-evolved with $\Gamma/J=8$,
thereby quenching across the quantum critical point at $\Gamma/J \approx 1.52$ \cite{Blote2002}. The results
obtained using a trajectory-resolved Weiss field (orange) 
improve upon those obtained by the trajectory-averaged Weiss field
(blue). For comparison, the results obtained using QuSpin's ODE Solver \cite{Weinberg2019} 
on a $5\times 5$ lattice are shown (dotted) in both panels; comparison results for a $7 \times 7$ lattice are currently beyond reach. The SDE
results are obtained from 5 batches of ${\mathcal N}= 10^5$ stochastic samples with a time-step $\Delta = 0.001$. The latter is chosen to accurately resolve the Loschmidt peaks. The faded lines indicate the standard error of the mean.}
\label{fig:losch}
\end{figure}

\begin{figure}[t]
\subfloat{\includegraphics[width =8.8cm]{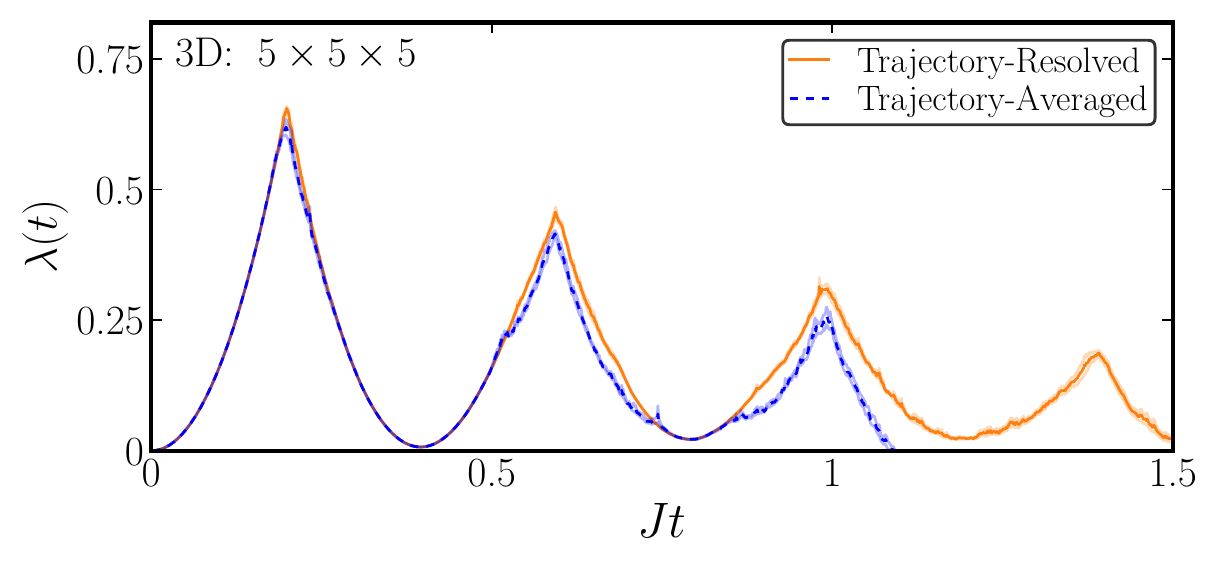}}
\caption{Time-evolution of the Loschmidt rate function $\lambda(t)$
following a quench in the three-dimensional quantum Ising model
on a $5\times 5 \times 5$ lattice. The spins
are initialized in the state $\ket{\psi(0)} = \frac{1}{\sqrt{2}}\Big( \ket{\Downarrow} + \ket{\Uparrow} \Big)$ corresponding to the superposition of degenerate ground states and time-evolved with $\Gamma/J=8$, thereby quenching across the quantum critical point at $\Gamma/J \approx 2.58$ \cite{Blote2002}.
The results
obtained using a trajectory-resolved Weiss field (orange) 
improve upon those obtained by the trajectory-averaged Weiss field
(blue). 
The SDE
results are obtained using 5 batches of ${\mathcal N}= 10^5$ stochastic samples and a time-step of $\Delta = 0.001$. The latter is chosen to resolve the Loschmidt peaks. The faded lines indicate the standard error of the mean.}
\label{fig:losch_125}
\end{figure}

 For simulations it is convenient to introduce a small time-delay $\delta$ into the trajectory-resolved Weiss field so that $m^z_j(t) = n_j^z(t-\delta)$. This enhances the numerical stability without approximation; see Appendix \ref{sec:numericalimpl}.
Throughout this work we solve the SDEs (\ref{eq:plus2})-(\ref{eq:zequat2}) using the Stratonovich-Heun predictor-corrector scheme.
 In simulations with a trajectory-resolved Weiss field we set $\delta = \Delta$, where $\Delta$ is the time-step, unless stated otherwise. In general, we find that a time-step of $\Delta = 0.01$ is sufficient for most of our simulations, unless the observable in question involves very small quantities, such as the Loschmidt amplitude. The use of a different time-step is highlighted in each instance.

\section{Simulations} \label{sec:simulations}
In this section we demonstrate the improvements for numerical
simulations when using trajectory-resolved Weiss fields. For
simplicity, we focus on quantum quenches of the nearest neighbor
quantum Ising model (\ref{eq:TFIM}) in both two and three dimensions, although improvements can also be seen in one-dimension. In
Fig. \ref{fig:losch} we show results for the Loschmidt rate function 
\begin{align}\lambda(t) = -\frac{1}{N}\ln |\braket{\psi(0)}{\psi(t)}|^2 \end{align}
following a quantum quench in two dimensions from an initial state with $\Gamma=0$ to
$\Gamma=8J$. Explicitly, we consider the initial state $\ket{\psi(0)} = \frac{1}{\sqrt{2}}\Big( \ket{\Downarrow} + \ket{\Uparrow} \Big)$, which is
the superposition of the symmetry broken groundstates: \begin{align}\ket{\Downarrow} = \prod_{i=1}^N \ket{\downarrow}_i, ~~~ \ket{\Uparrow} = \prod_{i=1}^N \ket{\uparrow}_i.\end{align} 
In Figs. \ref{fig:losch}\,(a) and \ref{fig:losch}\,(b) we show the results for a $5\times 5$ and a
$7 \times 7$ lattice respectively. It can be seen from Fig. \ref{fig:losch}\,(a) that
the trajectory-resolved approach allows simulations to be carried out
for longer time durations, before departures arise. For
comparison, we show results obtained by QuSpin's ODE solver \cite{Weinberg2019} for a $5\times 5$ lattice. The results are in excellent agreement until
$t\sim 3/J$, after which the stochastic fluctuations 
are not well sampled. 
Similar behavior is evident in Fig. \ref{fig:losch}\,(b), although we can only compare to the QuSpin results for the smaller system size, as $7\times 7$ is not possible at present.

In Fig. \ref{fig:losch_125} we show similar results for a 3D lattice with 125 sites  on a  $5 \times 5 \times 5$ grid. This is a far more challenging problem due to the increase in dimensonality. The results for the trajectory-resolved approach show clear Loschmidt peaks, which extend beyond those obtained by the trajectory-averaged approach. 
This complements earlier investigations \cite{Schmitt2018} which were unable to resolve the sharp non-analyticities due to significant finite-size effects.
\begin{figure}[t]
\subfloat{\centering
\includegraphics[width =8.8cm]{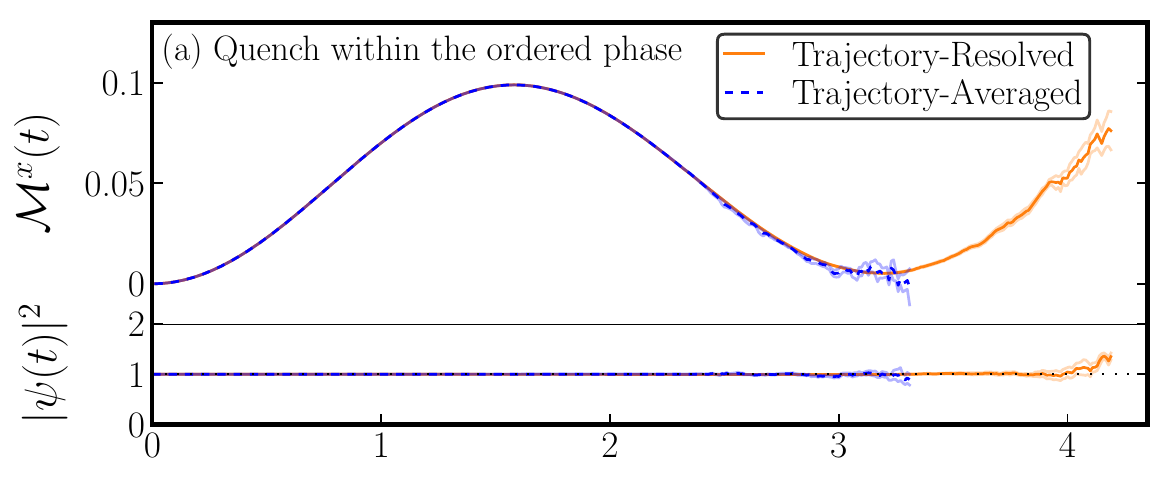}}

\vspace{-0.6cm}
\subfloat{\centering
\includegraphics[width =8.8cm]{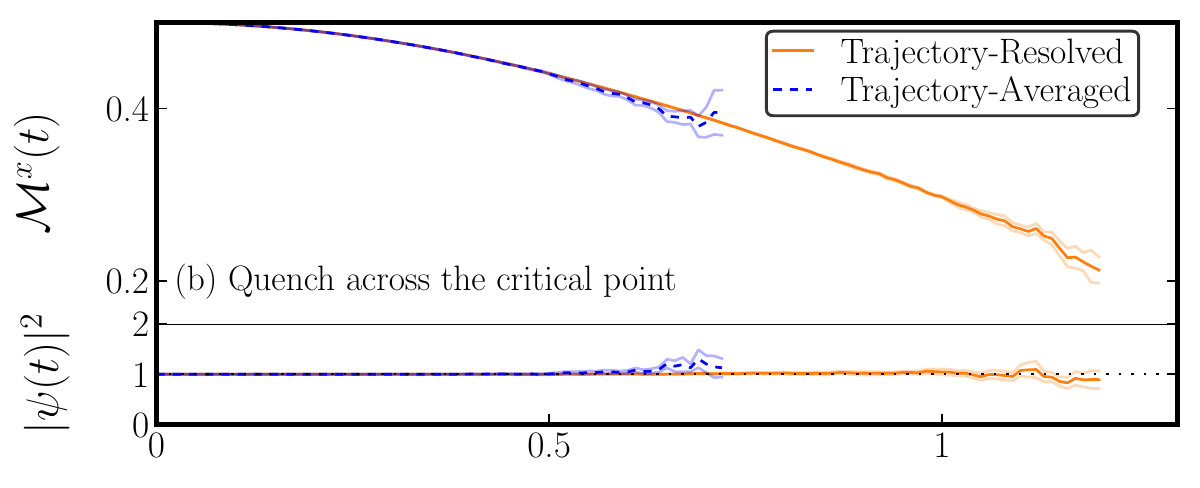}}

\vspace{-0.6cm}
\subfloat{\centering
\hspace{-0.08cm}
\includegraphics[width =8.775cm]{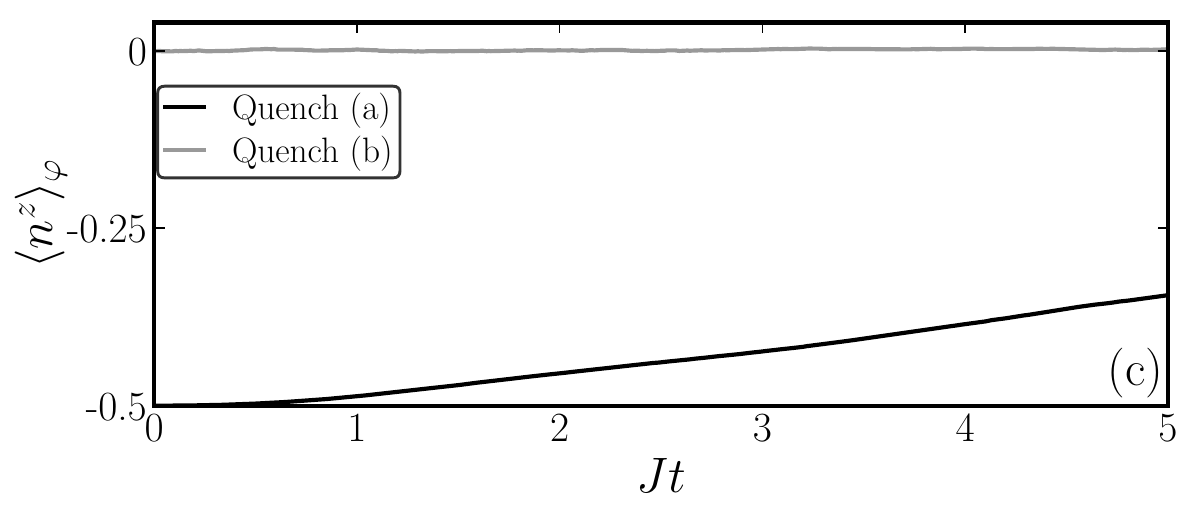}}
\caption{Time-evolution of the transverse magnetization $\mathcal{M}^x(t)$ following
quenches of the 2D quantum Ising model for a  $7 \times 7$ lattice with $49$ spins. (a) The
spins are initialized in the fully-polarized state along the $z$-axis
 $\ket{\Downarrow}$ and time-evolved with $\Gamma/J=0.2$. The results obtained with 
the trajectory-resolved Weiss field (orange) and the trajectory-averaged
Weiss field (blue) are in very good agreement. The normalization $|\psi(t)|^2$ indicates that the former reaches slightly longer
timescales. The success of the latter is consistent with the presence
of a non-vanishing trajectory-averaged Weiss field as shown in panel
(c). 
(b) The spins are prepared in the fully-polarized state $\ket{+}$ along the
$x$-axis and time-evolved with $\Gamma/J=0.1$. In this case, the
trajectory-resolved Weiss field performs much better than the
trajectory-averaged Weiss field. This is consistent with a vanishing
trajectory-averaged Weiss field as shown in panel (c). The results
are in very good agreement until the norm of the quantum state $|\psi(t)|^2$
departs from unity. The results in (a) correspond to 5 batches of $\mathcal{N} = 2\times 10^5$ stochastic samples while the results in (b) correspond to 5 batches of $\mathcal{N} = 6\times 10^5$ stochastic samples, all obtained with a time-step $\Delta = 0.01$. The faded lines indicate the standard error of the mean.}
\label{fig:magnet}
\end{figure}

\begin{figure}[t]
\includegraphics[width = 8.7cm]{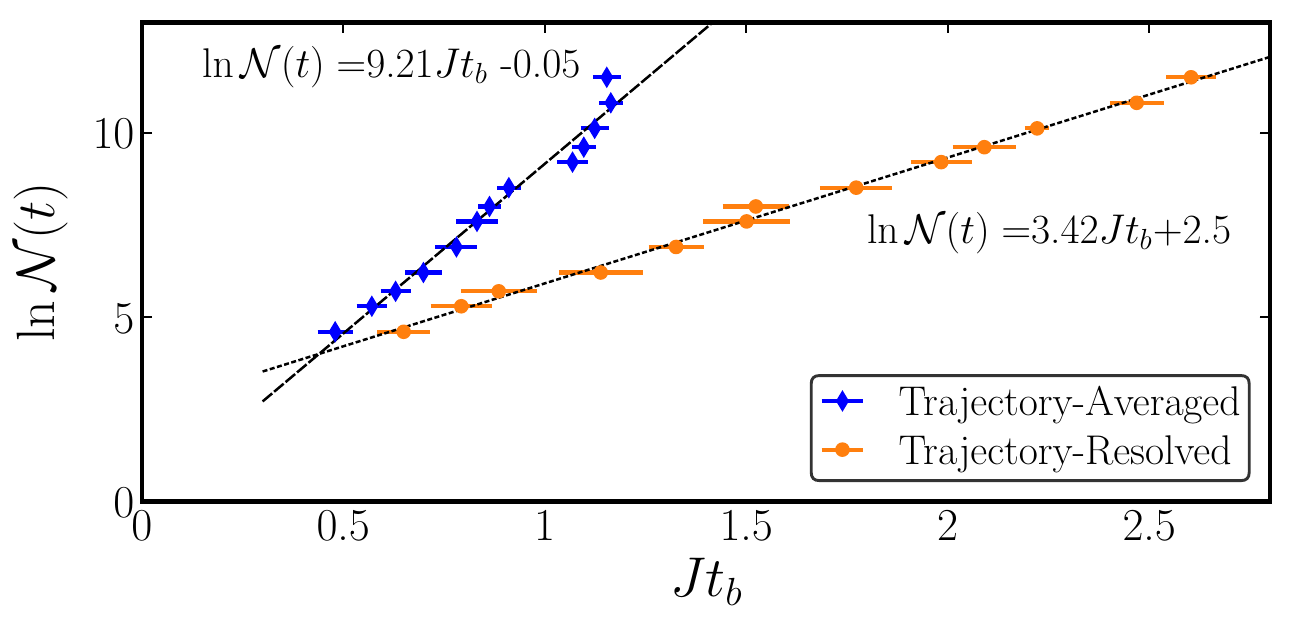}
\caption{Exponential scaling of the required number of samples $\mathcal
N$ versus the breakdown time $t_b$ of the simulations: $\mathcal{N} \sim c e^{\alpha t_b}$. The data
correspond to quantum quenches in the 2D quantum Ising model for a
$3\times 3$ array of spins. The spins are initialized in the
fully-polarized state $\ket{+}$ along the $x$-axis and are time-evolved
with $\Gamma/J=0.1$. The trajectory-resolved
Weiss field (circles) is shown to require less samples  than the trajectory-averaged Weiss field (diamonds) to reach a given time. This is confirmed by the coefficient $\alpha$ of the linear fit, which suggests a reduction in the exponent by a factor of approximately 2.7 for this simulation. Each of the data
points correspond to the mean of 10 batches of simulations of $\mathcal N$
runs with the standard error indicated by a bar.}
\label{fig:scalediagram}
\end{figure}

Having established the utility of trajectory-resolved Weiss fields, we
now examine the differences in performance for quenches that have
significant trajectory-averaged Weiss fields, and quenches that don't. In
Fig. \ref{fig:magnet}\,(a) we show results for the transverse magnetization $\mathcal{M}^x = \frac{1}{N}\sum_j \langle \hat{S}^x_j \rangle  $ following a
quantum quench within the ordered phase of the 2D quantum Ising model
(\ref{eq:TFIM}) on a $7 \times 7$ square lattice with $49$ sites.  We start from the fully-polarized state $\ket{\Downarrow}$ with all spins down, and time-evolve using the Hamiltonian with $\Gamma/J = 0.2$. As can be seen in Fig. \ref{fig:magnet}\,(c), this
quench is associated with a non-zero trajectory-averaged Weiss field over the duration of the simulation,
corresponding to a non-vanishing mean field in the initial state. It is evident from panel (a) that the trajectory-resolved Weiss fields perform better than the trajectory-averaged ones, although the
relative gains from their use is modest in this case. This can also be
seen from the behavior of the norm of the quantum state, $|\psi(t)|^2$, which
remains close to unity when the stochastic fluctuations are adequately
sampled \cite{Begg2019,Begg2020}. In the Appendices, we provide data for a quench in 3D within the ordered phase. The results demonstrate similar performance improvements when using the trajectory-resolved Weiss field.

 The difference between the two approaches is noticebly greater for quantum quenches without a significant trajectory-averaged Weiss field. This is illustrated in Fig. \ref{fig:magnet}\,(b) for a quantum quench
in the 2D quantum Ising model from the disordered phase to the ordered
phase. Specifically, the system is initialized in the fully-polarized
state in the x-direction, $\ket{+} = \prod_{j=1}^N \frac{1}{\sqrt{2}}(\ket{\downarrow} + \ket{\uparrow})_j$, and time-evolved with the Hamiltonian with $\Gamma/J = 0.1$. It can be seen that the simulation time with the
trajectory-resolved Weiss field is approximately double that of the
trajectory-averaged approach. For this particular quench the
stochastic trajectories rapidly spread out over the Bloch-sphere and the
trajectory-averaged Weiss field remains close to zero throughout the
evolution. As shown in Fig. \ref{fig:magnet}\,(c), the use of a trajectory-averaged Weiss field offers little advantage over the case with $m_j^z=0$. In contrast to the situation in equilibrium, where the utility of 
mean-fields increases with dimensionality, their use for dynamics is 
more subtle. In particular, the utility of Weiss fields can depend on 
the details of the quantum quench, and not just the dimensionality.

The utility of the trajectory-resolved approach can also be seen in
the scaling of the accessible timescale for numerical simulations with
the number of samples required. Following Refs. \cite{Begg2019,Begg2020} we define the breakdown time $t_b$ as the time when the normalization $|\psi(t)|^2$ differs from unity by 10\%. 
As discussed in Refs. \cite{Begg2019,DeNicola2019euclid,Begg2020}, the number of
samples $\mathcal N$ scales exponentially with the breakdown time, corresponding to the onset of strong fluctuations. Specifically, $\mathcal
N\sim c e^{\alpha t_b}$ where the growth exponent $\alpha$ depends on the details
of the quench. In Fig. \ref{fig:scalediagram} we compare the scaling of the
trajectory-averaged approach with the trajectory-resolved approach,
for a quench in the two-dimensional quantum Ising model with a
$3\times 3$ array of spins. It can be seen that the growth-exponent
$\alpha$ is significantly reduced for the trajectory-resolved case, in
comparison with the trajectory-averaged case. In Sections \ref{sec:stochasticnorm} and \ref{sec:growthfluct} we show that  this is related to a reduction in the fluctuations of the normalization of the stochastic state.

\begin{figure}[ht]
\begin{center}
\includegraphics[width = 8.7cm]{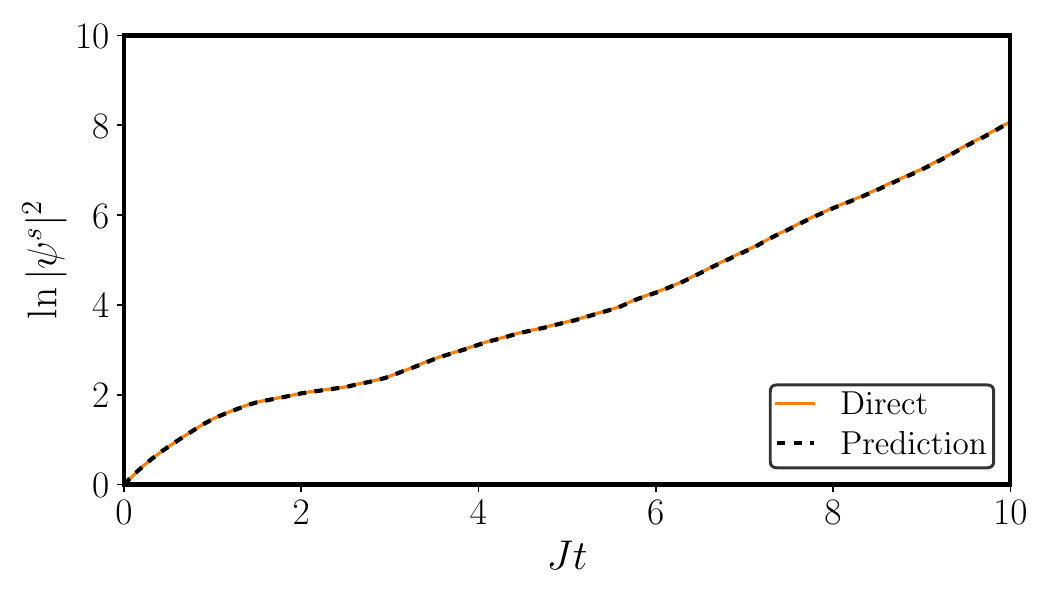}
\caption{
Growth of the stochastic state normalization $|\psi^s(t)|^2$ for a single trajectory following a quench of the 2D quantum Ising model with a $3 \times 3$ array of spins. 
The initial state $\ket{\psi(0)} = \ket{+}$ is evolved with $\Gamma/J= 0.3$. We compare a direct numerical evaluation of $|\psi^s(t)|^2$ (solid line) against the theoretical prediction (\ref{eq:normgrowth}) (dashed line). For the latter we use the value of $n_j^z(t)$ obtained from the numerical procedure as an input. 
The data is obtained using the Stratonovich-Heun scheme \cite{Ruemelin1982,Kloden1992} without time-delay and a time-step of $\Delta = 0.001$. The 
growth of the stochastic state normalization is required in order to 
maintain the overall normalization of the quantum state.}
\label{fig:lagSde}
\end{center}
\end{figure}

\section{Stochastic State Normalization} \label{sec:stochasticnorm}
In this section we discuss how the normalization of the stochastic state determines the sampling efficiency of the approach. 
To see this we note that an arbitrary normalized state  $\ket{\psi(t)}$ can be expressed as the average over normalized stochastic states, $|\ket{\psi^s(t)}$,  according to
\begin{align}
   \ket{\psi(t)}   = \Big\langle W(t) \, |\ket{\psi^s(t)} \Big\rangle_{\phi},
   \end{align} 
where $W(t) =  |\braket{\psi^s(t)}{\psi^s(t)}|^{1/2}$ is the norm of the stochastic state. It can be seen that $W(t)$ corresponds to the weight of each sample in the ensemble. A large spread directly inhibits sampling. As we show in Appendix \ref{sec:stochstateN}, the norm of the stochastic state grows monotonically in time. For the quantum Ising model this is given by 

\begin{align}
|\psi^s(t)|^2 &= \exp \left\{\int_0^{t} dt' ~  \sum_{j} \nu_j \left(\frac{1}{4} - n_j^z(t')^2 \right) \right\}, 
\label{eq:normgrowth}
\end{align} 
where $\nu_j = \sum_k |O_{jk}^{zz}|^2$; see Appendix \ref{sec:stochstateN}. 
We verify (\ref{eq:normgrowth}) in Fig. \ref{fig:lagSde} for a single stochastic trajectory following a quench in the 2D quantum Ising model. It can be seen from (\ref{eq:normgrowth}) that the normalization is controlled by deviations from the fully-polarized spin state. The growth of the norm with time is required in order to maintain the overall normalization of the quantum state, $|\psi(t)|^2 = 1$. To see this 
we note that 
\begin{align} |\psi(t)|^2=  \frac{1}{\mathcal{N}^2} \sum_{r,r'=1}^\mathcal{N}\braket{\psi_{r'}^{s}(t)}{\psi^{s}_r(t)}, \label{eq:normfails}\end{align}
where $r$ and $r'$ are independent sample indices. 
If $\ket{\psi^s_r(t)}$ was normalized then the overlaps in (\ref{eq:normfails}) would be less than or equal to unity. As such, $|\psi(t)|^2 < 1$. It follows that the normalization of  $\ket{\psi^s_r(t)}$ must grow with time in order to maintain the condition that $|\psi(t)|^2 = 1$. This reflects the independent decouplings of the forwards and backward time-evolution in the stochastic approach.

\begin{figure}[ht]
\subfloat{\includegraphics[width = 8.8cm]{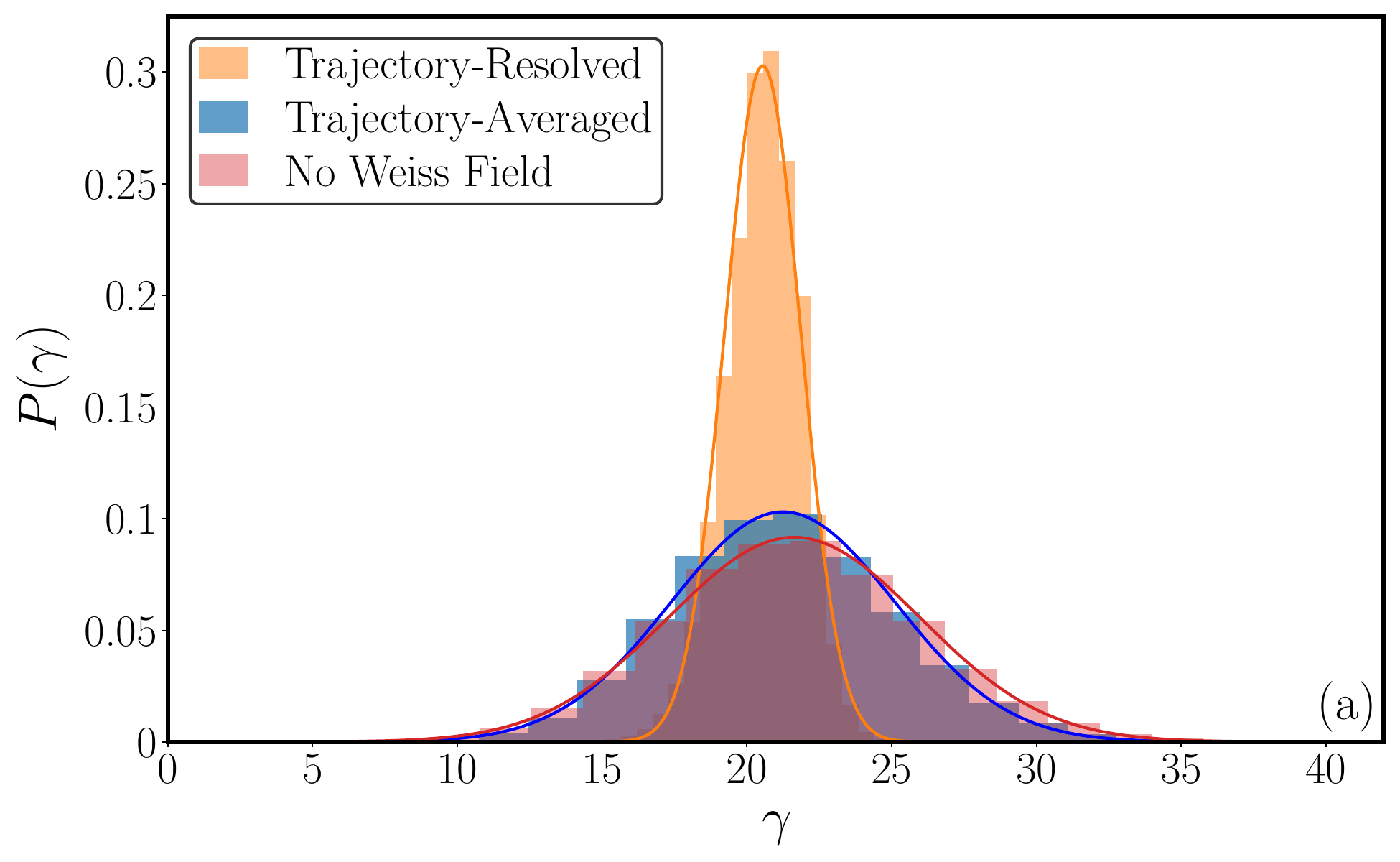}}

\subfloat{\includegraphics[width = 8.8cm]{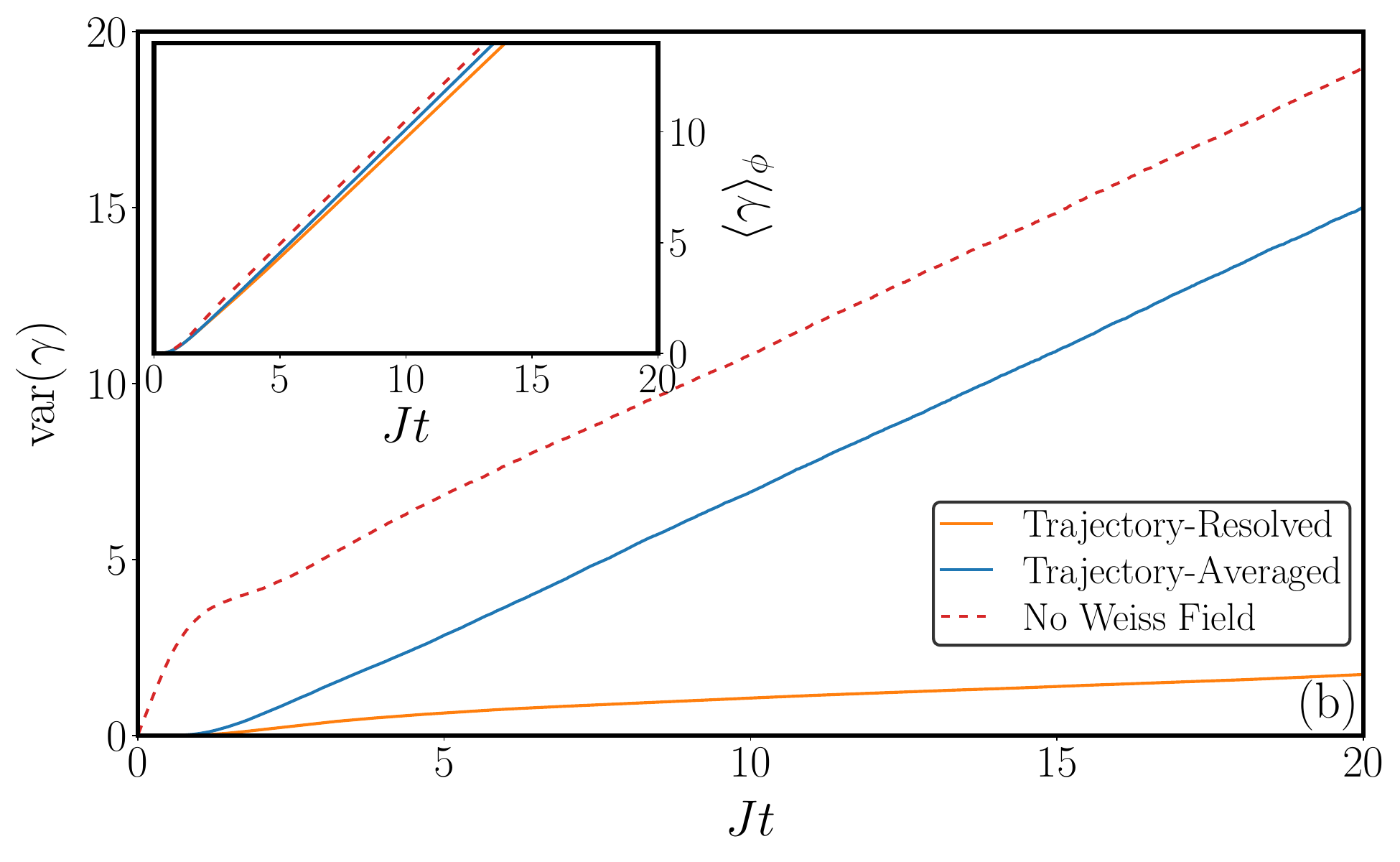}}
\caption{(a) Distribution $P(\gamma)$ at a fixed time $Jt = 20$ following a quench in the two-dimensional quantum Ising model for a $3 \times 3$ lattice of spins. We start in the fully-polarized state $\ket{\Downarrow}$ and quench to $\Gamma/J = 0.7$, taking $\mathcal{N} = 5 \times 10^4$ stochastic samples. The use of a trajectory-resolved Weiss field (orange) results in a narrower distribution than the trajectory-averaged case (blue) and the case without a Weiss field (red); this makes the dynamics easier to sample. The distributions are approximately normal, as indicated by the solid line fits. (b) Time-evolution of the variance $\text{var}(\gamma)$. The variance grows linearly following an initial transient. The growth rate for the trajectory-resolved case is lower than the other two cases, which again aids sampling. The variance for the trajectory-averaged case is only comparable to the trajectory-resolved case at very short timescales, when the state is well-approximated by fluctuations around a product state. Inset: time-evolution of the mean of the distribution $\langle \gamma \rangle_{\phi}$. The linear growth rate is similar for all three cases.} 
\label{fig:dist2}
\end{figure} 

\section{Growth of Fluctuations}\label{sec:growthfluct}
Having demonstrated that the norm of the stochastic state grows with time, we now examine its distribution. 
Given the exponential scaling of $W$ with time it is convenient to consider the distribution of $\gamma = \ln W$. In Fig. \ref{fig:dist2}\,(a) we show the distribution $P(\gamma)$ at a fixed time for different values of the Weiss field. It can be seen that $P(\gamma)$ is normally distributed, as indicated by the solid lines. 
The use of a trajectory-resolved Weiss field results in a narrower distribution than the other cases. In particular, it leads to a reduction in the extremal values  of $W = e^{\gamma}$ which contribute the most to ensemble averages. This leads to an improvement in the sampling efficiency. In Fig. \ref{fig:dist2}\,(b) we show the growth of the variance  $\text{var}(\gamma)$ as a function of time. 
It can be seen that the growth rate is reduced when using the trajectory-resolved Weiss field, even at late times. 
In contrast, the growth rate for the trajectory-averaged Weiss field eventually follows the case without a Weiss field; this was also observed in Ref. \cite{DeNicola2019euclid}, by sampling around a saddle-point trajectory in Euclidean time. 
In the inset of Fig. \ref{fig:dist2}\,(b) we show that the mean of the distribution $\langle \gamma \rangle_{\phi}$ changes very little with the choice of Weiss field. The improvements in the scaling are therefore attributed to the reduction of the variance with a trajectory-resolved Weiss field. Although the focus of this work is on 2D and 3D systems, similar improvements can also be seen in 1D, as shown in Appendix \ref{sec:additional_sims}.

In closing this section, we briefly comment on the role of fluctuations 
on the success of the trajectory-resolved Weiss field. Over timescales 
in which the dynamics can be approximated by fluctuations around a 
well-chosen product state trajectory, both the trajectory-averaged and 
the trajectory-resolved Weiss fields can efficiently encode the 
evolution. However, a key advantage of the trajectory-resolved approach, 
is that it can remain efficient beyond this timescale. In the 
trajectory-resolved approach, an entangled superposition such as a 
triplet state $\frac{1}{\sqrt{2}}(\ket{\uparrow \downarrow} + \ket{\downarrow \uparrow})$   can be obtained from two separate trajectories, $\ket{\uparrow \downarrow}$ and $\ket{\downarrow\uparrow}$,  
with each encoding its own Weiss field. In contrast, there is no single 
product state that approximates this triplet state. The 
trajectory-resolved approach therefore has a notable advantage for 
simulating quantum dynamics.

\section{Conclusion} \label{sec:conclusion}
In this work we have investigated the real-time dynamics of quantum
spin systems in two and three-dimensions by means of an exact
stochastic approach. We have shown that the use of a
trajectory-resolved Weiss field can significantly extend the
accessible simulation times, with an exponential improvement in the
sampling efficiency. We have illustrated the utility of this approach
for exploring dynamical quantum phase transitions in two and three
dimensions, although the applicability is broader. Our results address a critical shortage of exact
simulation techniques for non-equilibrium quantum spin systems in two
and three dimensions. It would be interesting to see if
trajectory-resolved Weiss fields can be used in other contexts, for
example in situations which traditionally involve expanding around a
single mean-field configuration.

\section{Acknowledgements}
SEB acknowledges support from the EPSRC CDT in Cross-Disciplinary
Approaches to Non-Equilibrium Systems (CANES) \textit{via} grant
number EP/L015854/1, as well as the support of the Young Scientist Training program at the Asia Pacific Center for Theoretical Physics (APCTP). MJB acknowledges support of the London Mathematical Laboratory. AGG acknowledges EPSRC grant EP/S005021/1. 
We are grateful to the UK Materials and Molecular
Modelling Hub for computational resources, which is partially funded
by EPSRC (EP/P020194/1 and EP/T022213/1). For the purpose of open access, the authors have applied a Creative Commons Attribution (CC BY) licence to any Author Accepted Manuscript version arising. The data supporting this article is openly available from the King's College London research data repository, KORDS, at https://doi.org/10.18742/24438967.


%

\appendix 

\section{Stochastic Hamiltonian} \label{sec:trajproof}
In this appendix we derive the stochastic Hamiltonian (\ref{eq:hstochwithgauge}) in the main text including the trajectory-resolved Weiss field. For simplicity, we consider Heisenberg models of the form 
\begin{align} 
\hat{H} = &   -  \sum_{ jk } \frac{1}{2} \Big( J^{xx}_{jk} \hat{S}^x_j \hat{S} ^x_{k}  + J^{yy}_{jk}  \hat{S}^y_j \hat{S} ^y_{k} + J^{zz}_{jk}  \hat{S}^z_j \hat{S} ^z_{k} )  \nonumber \\ & - \sum_{j} \Big( h_j^x \hat{S}^x_j + h_j^y \hat{S}^y_j + h_j^z \hat{S}^z_j \big), \label{eq:xxzham} 
\end{align}
where $J^{aa}_{jk}$ is the exchange interaction in direction $a$ and $h_j^a$ is an applied magnetic field. 
\subsection{Ito Convention}\label{sec:ito} 
As discussed in the main text, the stochastic Hamiltonian (\ref{eq:hstochwithgauge}) is obtained via the change of variables 
\begin{align}
& \varphi^{'a}_{j}(t)  = \varphi^{a}_{j}(t) + \sqrt{i} \sum_k J^{aa}_{jk} m^a_k(t). \label{eq:transagain}
\end{align}
Here we consider the specific case of the trajectory-resolved Weiss field for which $m^a_k(t) = n_k^a(t)$, as given in (\ref{eq:nz}). Substituting (\ref{eq:transagain}) into Eq. (\ref{eq:HStransf}) and re-arranging the terms generates the stochastic Hamiltonian (\ref{eq:hstochwithgauge}). However, in principle one should also check for the possibility of a non-trivial Jacobian matrix $\bm{\mathcal{J}}$ associated with the field transformation (\ref{eq:transagain}): \begin{align} \mathcal{J}^{ab}_{j'j}(t',t) = \frac{\delta \varphi^{'a}_{j'}(t')}{\delta \varphi^b_{j}(t)}. \label{eq:jacobmat}\end{align} Here we show that the Jacobian $ \mathcal{J} = \text{det}(\bm{\mathcal{J}})$ is in fact trivial. To see this, we note that  
since $n_j^a(t)$ does not depend on future noise configurations the matrix (\ref{eq:jacobmat}) is lower-triangular in the time domain. The Jacobian therefore reduces to a product of equal-time contributions, which lie along the diagonal of the matrix. To exploit this, we choose the following discrete time version of (\ref{eq:transagain}) including the trajectory-resolved Weiss field: 
\begin{align}
 \varphi^{'a}_{j\tau} &  = \varphi^{a}_{j\tau} + \sqrt{i} \sum_k J^{aa}_{jk} n^a_{k(\tau-1)} \label{eq:discretetrans}.
\end{align}
Here we employ the  discrete real-time index $\tau$, which should not be confused with Euclidean time. 
Since $n^a_{k(\tau-1)}$ does not depend on $\varphi^{a}_{j\tau}$, the diagonal entries of the matrix (\ref{eq:jacobmat})  are unity.  As such the Jacobian is unity. In the language of stochastic processes this aligns with the Ito definition since the fields $\varphi^a_{j\tau}$ and $n^a_{j(\tau-1)}$ are uncorrelated; the field $\varphi^a_{j\tau}$ is associated with the stochastic evolution operator $\hat{U}^s(\tau-1,\tau)$. 
In the next section we derive the stochastic Hamiltonian in the Stratonovich formalism. 

\subsection{Stratonovich Convention}\label{sec:strat}
For numerical simulations it is often convenient to work in the Stratonovich formalism due to the robustness of the associated numerical schemes. 
Here we consider models of the form (\ref{eq:xxzham}) with $J_{jj}^{aa} = 0$. 
In the absence of a Weiss field  the Stratonovich and Ito SDEs (\ref{eq:SDEs}) coincide \cite{DeNicola2019long}. The same is true in the presence of a trajectory-averaged Weiss field. However, new terms arise in the Stratonovich SDEs with a trajectory-resolved Weiss field. To see this we first consider the evolution equation
\begin{align}
i \partial_t \ket{\psi^s(t)} = \hat{H}^s(t) \ket{\psi^s(t)} \label{eq:stochschro},
\end{align}
 where $\hat{H}^s(t)$ is the stochastic Hamiltonian (\ref{eq:hstochwithgauge}) in the Ito form. 
Since $\hat{H}^s(t)$ is non-interacting, the evolution equation on site $j$ can be written as 
\begin{align}
 d \ket{\psi^s_j(t)} = \mathcal{A}_j dt + \sum_{ka} \mathcal{B}^a_{jk} dW^a_k , 
\end{align}
where $W^a_k$ are independent Weiner processes associated with the white noises $\phi_k^a$. Here
\begin{align} &  \mathcal{A}_j= \\ & \Big(  i \sum_{a}  h_j^a \hat{S}^a_{j}  + i \sum_{ka} J^{aa}_{jk} n_k^a \hat{S}^a_{j}  - \frac{i}{2} \sum_{ka}  J_{jk}^{aa}n^{a}_{j}n^{a}_{k} \hat{\mathbb{I}} \Big) \ket{\psi^s_j(t)}, \nonumber \\  
& \mathcal{B}^a_{jk} =  \Big(   \sqrt{i}  \big(\hat{S}^a_j - n_j^a \hat{\mathbb{I}} \big)  O^{aa}_{jk}\Big)\ket{\psi^s_j(t)}, 
\end{align}
where $O^{aa}_{jk}$ is defined in Section \ref{sec:stochasticformalism} of the main text.
The associated Stratonovich equations are given by 
\begin{align}
 d \ket{\psi^s_j(t)} = & \mathcal{A}_j- \frac{1}{2}\sum_{kla} \mathcal{B}^a_{lk}\frac{\partial \mathcal{B}^a_{jk}}{d \lambda_l} + \sum_{ka} \mathcal{B}^a_{jk} dW^a_k , \label{eq:conversionformula} 
\end{align}
where $\lambda_l \equiv \{ \ket{\psi^s_l}, \bra{\psi^{s}_l} \}$ \cite{Kloden1992}. 
Comparison with (\ref{eq:stochschro}) allows one to define the stochastic Hamiltonian in the Stratonovich formalism 
 \begin{align} & \hat{H}^{s}(t)  = \label{eq:hstochwithtraj}\\& - \sum_{ja}  h_j^a \hat{S}^a_{j} - \sum_{jka} J^{aa}_{jk} n_k^b \hat{S}^a_{j} \nonumber  - \frac{1}{\sqrt{i}} \sum_{ja}  \varphi_j^a   \big(\hat{S}^a_j - n_j^a \hat{\mathbb{I}} \big)\\ & + \frac{1}{2} \sum_{jka}  J_{jk}^{aa}n^{a}_{j}n^{a}_{k} \hat{\mathbb{I}}+ i \sum_{ja} \frac{\nu^a_j}{2} \Big( \frac{1}{4}     -     n^a_{j}(t)^2\Big)\hat{\mathbb{I}} , \nonumber \end{align}
where $\nu^a_j = \sum_k |O_{jk}^{aa}|^2$.
 Only the final term differs from the Ito form of the  stochastic Hamiltonian given in (\ref{eq:hstochwithgauge}). In the next section, we will show that this contribution determines the normalization of the time-evolving stochastic state.  

\subsection{Stochastic State Normalization}\label{sec:stochstateN} 
We consider the normalization of the stochastic state following an infinitesimal time interval $\delta t$: 
\begin{align}
& \braket{\psi^s(t+\delta t)}{\psi^s(t+\delta t)} = \bra{\psi^s(t)} e^{i \hat{H}^{s\dagger} \delta t}e^{ -i \hat{H}^s \delta t }\ket{\psi^s(t)} \nonumber \\ 
& = \braket{\psi^s(t)}{\psi^s(t)} + \delta t \Big(-i \langle \hat{H}^s \rangle  +  i \langle \hat{H}^{s\dagger} \rangle \Big)+  \label{eq:normexp}\\&  \delta t^2  \Big(- \frac{1}{2}\langle  (\hat{H}^s)^2 \rangle   - \frac{1}{2}\langle ( \hat{H}^{s\dagger})^2 \rangle + \langle  \hat{H}^s \hat{H}^{s\dagger}\rangle \Big) + \mathcal{O}(\delta t^3) \nonumber. 
\end{align}
In the Stratonovich formalism the $\mathcal{O}(\delta t^2)$ contributions vanish as $\delta t \rightarrow 0$ and we therefore consider the $\mathcal{O}(\delta t)$ term. The Hermitian terms in (\ref{eq:hstochwithtraj}) cancel out at this order, leaving just the non-Hermitian part 
\begin{align} \label{eq:nhterm}  - \frac{1}{\sqrt{i}} \sum_{j}  \varphi_{j}^a(t) &  \big(\hat{S}^a_j - n_j^a(t) \hat{\mathbb{I}} \big) \\ &+  i \sum_{ja} \frac{\nu^a_j}{2} \Big( \frac{1}{4}     -     n^a_{j}(t)^2\Big)\hat{\mathbb{I}}. \nonumber \end{align}
 The expectation value of the first term in (\ref{eq:nhterm}) vanishes, leaving only the second term to contribute to (\ref{eq:normexp}). Integrating over time yields 
\begin{align}
|\psi^s(t)|^2 &= \exp \Big\{\int_0^{t} dt' ~  \sum_{ja} \nu^a_j \Big(\frac{1}{4} - n_j^a(t')^2 \Big)\Big\}.  \label{eq:normgrowthappen}
\end{align}
The result (\ref{eq:normgrowthappen}) can also be obtained in the Ito formalism using (\ref{eq:hstochwithgauge}). In this case the contributing term appears at second order in the expansion (\ref{eq:normexp}) due to the properties of Weiner differentials in the Ito description. Specifically, the expectation value of their square does not vanish and one should use $d W_k d W_k' \rightarrow dt~ \delta_{k k'}$ in (\ref{eq:normexp}) \cite{Kloden1992}. This makes some of the second order terms in (\ref{eq:normexp}) first order. In contrast, $dW_k d W_{k'} \rightarrow 0$ in the Stratonovich description.  
The coefficient $\nu^a_j = \sum_k |O_{jk}^{aa}|^2$  in (\ref{eq:normgrowthappen}) can be interpreted as the effective interaction strength at site $j$ in the $a$-direction. This can be seen by using the bond noise description introduced in Ref. \cite{Begg2020}. In this description, $\nu^a_j = J^{a}Z^a_j$, where $Z_j^a$ is the number of interactions experienced by spin $j$ in the $a$-direction; here, for simplicity, we assume that the interactions are of equal strength $J^a.$

\section{Time-delay Formalism}\label{sec:numericalimpl}
In previous works \cite{Begg2019,Begg2020,DeNicola2021} the Stratonovich-Heun scheme \cite{Ruemelin1982,Kloden1992} has been successfully used to integrate the SDEs (\ref{eq:SDEs}). In the presence of a trajectory-resolved Weiss field the additional term appearing in (\ref{eq:hstochwithtraj}) necessitates the use of smaller time-steps.
To circumvent this, we introduce a time-delay $\delta$ into the definition of the trajectory-resolved Weiss field, such that $m^z_j(t) =  n_j^z(t- \delta) $: 
\begin{align}
 \varphi^{'a}_{j\tau} &  = \varphi^a_{j\tau} + \sqrt{i} \sum_k J^{aa}_{jk} n^{a}_{j(\tau-1 - \delta)} \label{eq:discretetrans2}.
\end{align}
That is to say, the Weiss field is determined by the spin configuration at a slightly earlier time-step.
This results in stochastic delay differential equations (SDDEs) which are easier to implement numerically. 
The Ito-Stratonovich conversion formula for SDDEs is identical to the undelayed case \cite{GuillouzicThesis}. In this approach, the last term in (\ref{eq:hstochwithtraj}) no longer appears.
As such, the Ito and Stratonovich SDDEs coincide. 
A small delay $\delta$ provides numerical stability at large time-steps, as afforded by other Stratonovich schemes.
 For times  $t<\delta$ we set $m^z(t-\delta)$ equal to the initial magnetization. 
\begin{figure}[t]
\subfloat{\includegraphics[width = 8.7cm]{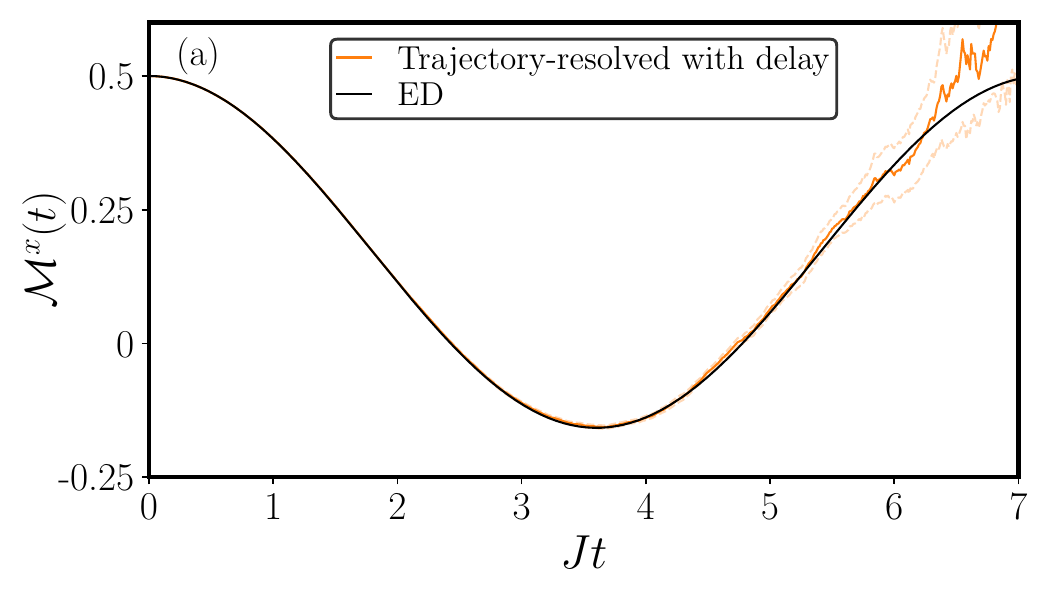}}

\vspace{-0.5cm}
\subfloat{\includegraphics[width = 8.7cm]{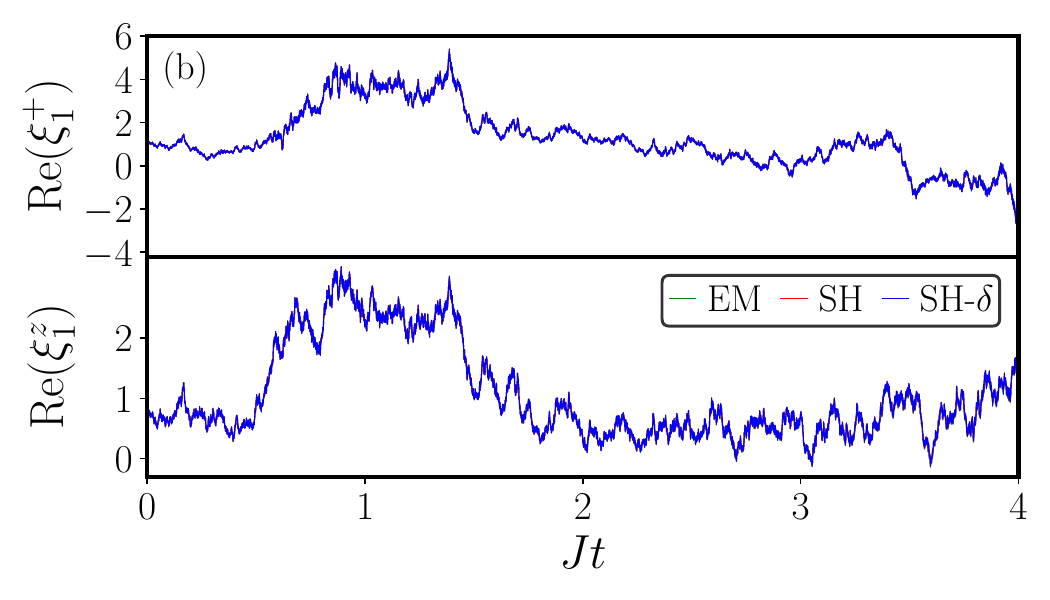}}
\caption{(a) Transverse magnetization $\mathcal{M}^x(t)$ following a quench of the quantum Ising model from  $\prod_i^N\ket{+}_i$ to $\Gamma/J = 0.3$ with $N =3$. The data correspond to a trajectory-resolved Weiss-field $m_i^z = n_i^z(t-\delta)$ with a large delay $\delta = 0.2$ (orange). The results are in excellent agreement with exact diagonalization (ED) until the breakdown time $t_b$ associated with finite sampling.  
We use 10 batches of $\mathcal{N} = 10^5$ samples, and a time-step of $\Delta = 0.01$. 
The solid orange line corresponds to the mean and the light orange lines indicate the standard error. 
(b) Time-evolution of the real parts of $\xi^+_1$ and $\xi^z_1$ for a single trajectory using the Ito Euler-Maruyama (EM), Stratonovich-Heun (SH) and delayed Stratonovich-Heun (SH-$\delta$) schemes. We consider the set-up as (a) but with $\Gamma/J = 0.1$ and a small time-step of $\Delta = 10^{-5}$. The results for the different integration schemes are almost indistinguishable, which demonstrates that the time-delayed result converges to the undelayed result in the limit of a small delay $\delta$.
In this simulation we use complex noises residing on each bond; see \cite{Begg2020} for a discussion. 
}
\label{fig:arblag}
\end{figure}

To perform the numerical integration of (\ref{eq:plus2}) and (\ref{eq:zequat2}) for the quantum Ising model in the presence of a delayed trajectory-resolved Weiss field we employ the predictor-corrector scheme used in  Ref. \cite{Cao2015}. For clarity, we  re-state the SDEs from the main text in the presence of a trajectory-resolved Weiss field with delay: 
\begin{align}
-i  \dot{\xi}^+_j(t) =& \frac{\Gamma}{2} (1  - \xi^{+^2}_j(t)) + \\&  \xi^+_j(t)\Big( \sum_{k} J_{jk}^{zz}n_k^z(t-\delta)  + \frac{1}{\sqrt{i}} \sum_{k}  O^{zz}_{jk} \phi^z_k(t)\Big) \nonumber 
\end{align}  
\begin{align}
& -i \dot{\xi}^z_j(t) = -  \Gamma \xi^+_j(t)   +   \sum_{k} J_{jk}^{zz}n_k^z(t-\delta)  +  \\ & \sum_{k}  J_{jk}^{zz}n^{z}_{j}(t-\delta)n^{z}_{k}(t-\delta) +\frac{1}{\sqrt{i}} \sum_{k}O^{zz}_{jk} \phi^z_k(t)\big(1 + 2n_j^z(t-\delta)\big) \nonumber
\end{align}                    
These can be written in the canonical form
\begin{align} d\xi^{a}_j = A^{a}_j(\xi(t),\xi(t-\delta)) dt + \sum_{k} B^{a}_{jk}(\xi(t),\xi(t-\delta))dW_{k}^a,\end{align}
where 
\begin{align}
& A^+_j\big(\xi(t),\xi(t-\delta)\big) = \frac{i\Gamma}{2} \big(1  - \xi^{+^2}_j(t)\big) \nonumber\\& ~~~~~~~~~~~~~~~~~~~~~~~~~ + i \xi^+_j(t)  \sum_{k} J_{jk}^{zz}n_k^z(t-\delta)  ,\\ 
& B^+_{jk}\big(\xi(t),\xi(t-\delta)\big) =   \sqrt{i} \xi^+_j(t) O^{zz}_{jk},\end{align}\begin{align} 
& A_j^z\big(\xi(t),\xi(t-\delta)\big) = - i \Gamma \xi^+_j(t) +i \sum_{k} J_{jk}^{zz}n_k^z(t-\delta)   , \nonumber \\ &  ~~~~~~~~~~~~~~~~~~~~~~~~~ + i\sum_{k}  J_{jk}^{zz}n^{z}_{j}(t-\delta)n^{z}_{k}(t-\delta), \\  
& B^z_{jk}\big(\xi(t),\xi(t-\delta)\big) = \sqrt{i} O^{zz}_{jk} \big(1 + 2n_j^z(t-\delta)\big). 
\end{align}
Moving to a discrete time index, the numerical update scheme \cite{Cao2015} is given by 
\begin{align} &\xi_{j\tau+1}^{a}  = \xi_{j\tau}^{a} + \frac{ \Delta }{2}\big(A^{a}_j(\xi_\tau,\xi_{\tau-\delta}) + A^{a}_j(\tilde{\xi}_{\tau+1},\xi_{\tau-\delta+1})\big) \nonumber \\ &+ \frac{1}{2}\sum_k\big(B^{a}_{jk}(\xi_\tau,\xi_{\tau-\delta}) + B^{a}_{j k}(\tilde{\xi}_{\tau+1},\xi_{\tau-\delta+1}\big)\Delta W^a_{k_\tau}, \end{align}
where $\delta$ is the time-delay and $\Delta$ is the time-step. The prediction step is given by 
\begin{equation}
\tilde{\xi}^{a}_{j\tau+1} = \xi^{a}_{j\tau} + \Delta A^{a}_j(\xi_\tau,\xi_{\tau-\delta})  + \sum_k B^{a}_{jk}(\xi_\tau,\xi_{\tau-\delta})  \Delta W^a_{k\tau}.
\end{equation}
In the absence of a time delay $\delta$ this coincides with the Stratonovich-Heun predictor-corrector scheme \cite{Ruemelin1982,Kloden1992}. We therefore refer to this as the delayed Stratonovich-Heun scheme.  
In order to illustrate the viability of this method, in Fig. \ref{fig:arblag}\,(a) we show results obtained with a relatively large time-delay $\delta = 0.2$ and a time-step of $\Delta= 0.01$. The results are in very good agreement with exact diagonalization (ED) until the breakdown time $t_b$. 
In Fig. \ref{fig:arblag}\,(b) we show the time-evolution of the stochastic variables $\xi^+_1(t)$ and $\xi^z_1(t)$ following a quantum quench. It can be seen that the results obtained via the Ito Euler-Maruyama scheme, the Stratonovich-Heun scheme and the delayed Stratonovich-Heun scheme with $\delta = \Delta$ are in excellent agreement. In practice, this is only true for a sufficiently small time-step. Nonetheless, we find that results obtained in the delay formalism are robust at large time-steps. This is evident from the simulations presented in the main text.

\section{Gauge-P Formalism} \label{sec:driftgauges}
In this section we use the gauge-P phase space formalism \cite{Deuar2002} to give an alternative derivation of the SDEs (\ref{eq:plus2}) and (\ref{eq:zequat2}) for the specific case of the quantum Ising model. We follow the discussion in Appendices B-D of our recent work \cite{Begg2020}, and include a trajectory-resolved Weiss field. 
We begin with a decomposition of the density matrix $\hat{\rho}$ in terms of coherent states $\ket{\lambda_j}$: 
\begin{align}
\hat{\rho} = \int d^2{\lambda}d^2\lambda'd^2\omega ~ P(\lambda,\lambda',\omega) ~ e^\omega \prod_j \ket{\lambda_j}\bra{\lambda_j'}
\label{eq:decomppos}, \end{align}  
where  $\lambda,\lambda' \in \mathbb{C}$ and $\omega$ is a complex weight. 
The decomposition (\ref{eq:decomppos}) is not unique due to the over-completeness of the coherent-state basis; see for example \cite{Deuar2002,DeuarThesis}.  Substituting \ref{eq:decomppos} into the Liouville equation for $\hat{\rho}$ yields a Fokker--Planck equation provided boundary terms vanish \cite{Drummond1980}.
 This in turn yields SDEs for the variables $\lambda, \lambda', \omega$. It is possible to alter these equations and move between different representations of $P(\lambda,\lambda',\omega)$. In the Ito formulation, the SDEs can be written in the form 
\begin{align}  
&  \dot{\uplambda}_{j} = A_{j} - \sum_{k } g_{k}B_{j k} +  \sum_k B_{j k} \phi_k,\label{eq:langevinfinal} \\ &
\dot{\omega} = V - \frac{1}{2} \sum_k g_{k}^2 + \sum_k g_{k}
\phi_k. \label{eq:langevinfinal2}
\end{align} 
where $A_j(\lambda)$ is the drift term, $B_{jk}(\lambda)$ is the noise-matrix and $\uplambda = \{\lambda,\lambda'\}$. The term $V(\uplambda)$ is needed to ensure a valid Fokker--Planck equation. 
The coefficients $g_k(\uplambda)$ are arbitrary functions known as drift gauges \cite{Deuar2002,DeuarThesis}; these are introduced by adding trivial terms to the Liouville equation. 
 The drift gauges re-weight trajectories via their influence on $\omega$ and $\uplambda$.

For the quantum Ising model we employ SU(2) spin coherent states $\ket{z_j} = \exp \left(e^{z_j} \hat{S}^+_j\right)\ket{\downarrow}$ \cite{Barry2008,Ng2013}\cite{Begg2020}, and their weighted counterparts $\ket{z,\omega} =  \prod_j e^{\omega_j}\ket{z_j}$.
Since the forward and backwards time-evolution protocols are independent in our approach, we consider a decomposition over states rather than density matrices \cite{Begg2020}: 
\begin{align} \ket{\psi} = \int d^2z d^2\omega ~P(z,\omega)~ \ket{z,\omega}. \label{eq:waveansatz}\end{align}
With this parametrization (\ref{eq:langevinfinal}) and (\ref{eq:langevinfinal2}) become 
\begin{align}
& - i\dot{z}_j =   -\Gamma \sinh(z_j) - \frac{1}{2}\sum_{l}J_{jl} + \frac{1}{\sqrt{i}}\sum_k  O^{zz}_{jk} (\phi_k-g_k), \label{eq:zi}\\ &
-i\dot{\omega} = \sum_j 
\frac{\Gamma }{2}e^{z_j} + \sum_{jl} \frac{1}{8}J_{jl}  + \frac{i}{2} \sum_k g_{k}^2 - i \sum_k g_{k}\phi_k  \label{eq:omegai},
\end{align}
where $\omega = \sum_j \omega_j$ and $\bm{O}$ is defined in Section \ref{sec:stochasticformalism} of the main text. The details of these calculations are presented in Ref. \cite{Begg2020}.
The Weiss fields $m_j^z$ can be introduced by setting
\begin{align} g_k(z) =
  - \sqrt{i}\sum_j\left(\frac{1}{2} + m_j^z\right)O^{zz}_{jk}. \label{eq:isinggaugechoice} 
\end{align}
 In our previous work \cite{Begg2020} $m_j^z$ is a time-dependent parameter which is independent of $z$ and $z^*$. Here, we allow it to depend explicitly on these parameters. The trajectory-resolved Weiss field corresponds to $m_j^z(z,z^*) = n^z_j(z,z^*).$
In these notations, the Ito SDEs are given by
\begin{align}
 - i\dot{z}_j =& 
 \sum_l J_{jl} m_l^{z} +  \frac{1}{\sqrt{i}}\sum_k  O^{zz}_{jk} \phi_k  -\Gamma \sinh(z_j) , \label{eq:zi2}\\ 
-i\dot{\omega}_j = &  
\frac{\Gamma }{2}e^{z_j}  - \frac{1}{2}\sum_l J_{jl} m_l^{z} - \frac{1}{\sqrt{i}}\Big(\frac{1}{2}+m_j \Big)\sum_k  O^{zz}_{jk} \phi_k   \nonumber \\ &   - \frac{1}{2} \sum_{k}  J_{kj}m^{z}_{k}m^{z}_{j}. \label{eq:omegai3}
\end{align}	
Setting $\xi^+_j = \ln z_j$ and using Ito's lemma, (\ref{eq:zi2}) and (\ref{eq:omegai3}) coincide with (\ref{eq:plus2}) and (\ref{eq:zequat2}) for the quantum Ising model. 
In the case of no Weiss field, or a trajectory-averaged Weiss field, the Stratonovich and Ito SDEs (\ref{eq:zi2}) and (\ref{eq:omegai3}) are the same. For the trajectory-resolved Weiss field the Stratonovich SDEs contain  an additional term 
$-i \frac{\nu^z_j}{2} \big(\frac{1}{4} - n_j^z(z,z^*)^2\big)  \label{eq:shiftterm} $ which should be added to the right-hand side of (\ref{eq:omegai3}), in conformity with (\ref{eq:hstochwithtraj}). As discussed in Section (\ref{sec:stochstateN}) this is the only term that contributes to the normalization of the stochastic state.
For further details on the gauge-P approach, including the use of time-delays, see Ref. \cite{BeggThesis}.
\begin{figure}[t]
\subfloat{\centering
\includegraphics[width =8.8cm]{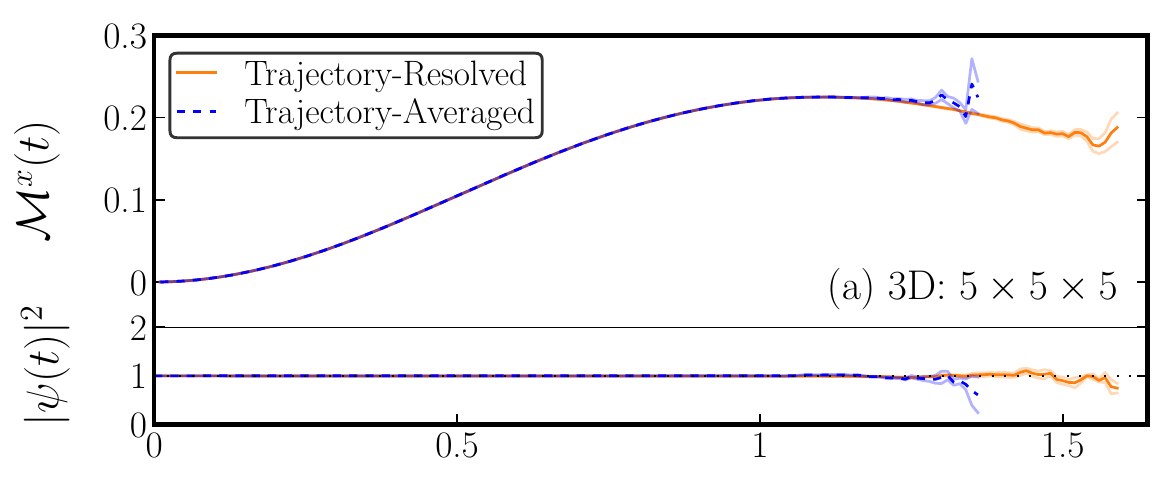}
}

\vspace{-0.4cm}
\subfloat{\centering
\includegraphics[width =8.8cm]{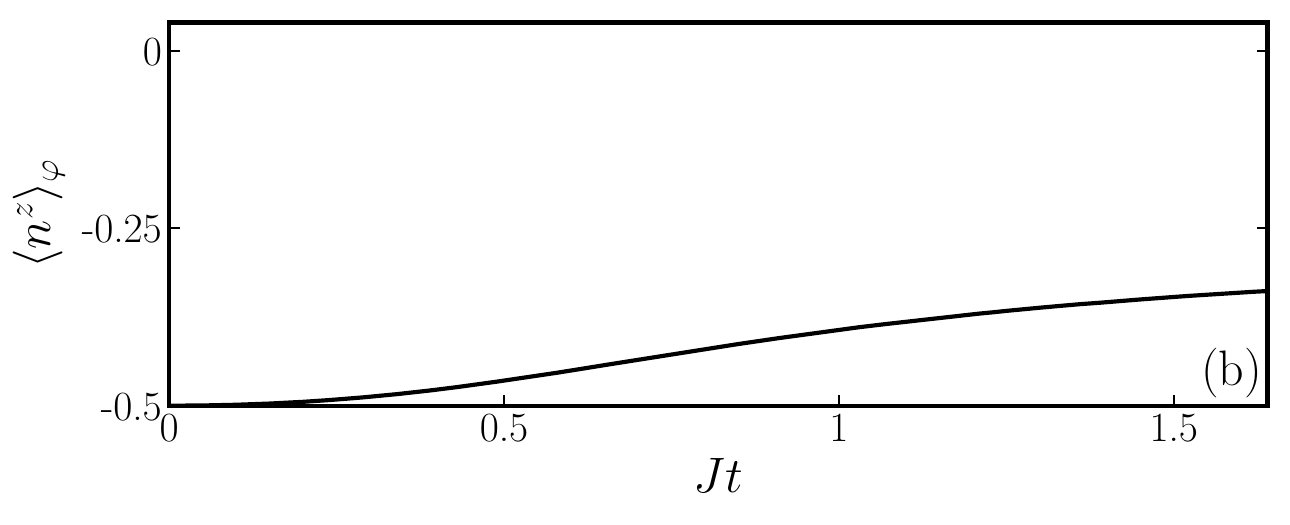}
}
\caption{(a) Time-evolution of the transverse magnetization $\mathcal{M}^x(t)$ following
quenches in the 3D quantum Ising model. (a)
Results for a $5\times 5 \times 5$ lattice with $125$ spins. The
spins are initialized in the fully-polarized state along the $z$-axis
 $\ket{\Downarrow}$ and time-evolved with $\Gamma/J=0.7$. The results obtained with
the trajectory-resolved Weiss field (orange) and the trajectory-averaged
Weiss field (blue) are in very good agreement. The normalization $|\psi(t)|^2$ indicates that the former reaches slightly longer
timescales. 
The results correspond to 4 batches of $\mathcal{N} = 10^5$ stochastic samples, 
obtained using a time-step of $\Delta = 0.01$. The faded lines indicate the standard error of the mean. (b) Trajectory-averaged Weiss field vs time, showing a non-vanishing trajectory-averaged Weiss field .}
\label{fig:magnet_3D}
\end{figure}

\section{Additional Simulations}\label{sec:additional_sims}
In this section, we provide numerical simulations examples in other spatial dimensions to complement those in the main text. 
In Fig. \ref{fig:magnet_3D}, we show results for a quantum quench in the ordered phase of the 3D quantum Ising model. It can be seen that trajectory-averaged Weiss field extends the simulation time in a manner akin to that of Fig. 4(a), due to the presence of a significant trajectory-averaged Weiss field for a quench in the ordered phase. 

\begin{figure}[t]
\subfloat{\includegraphics[width = 8.8cm]{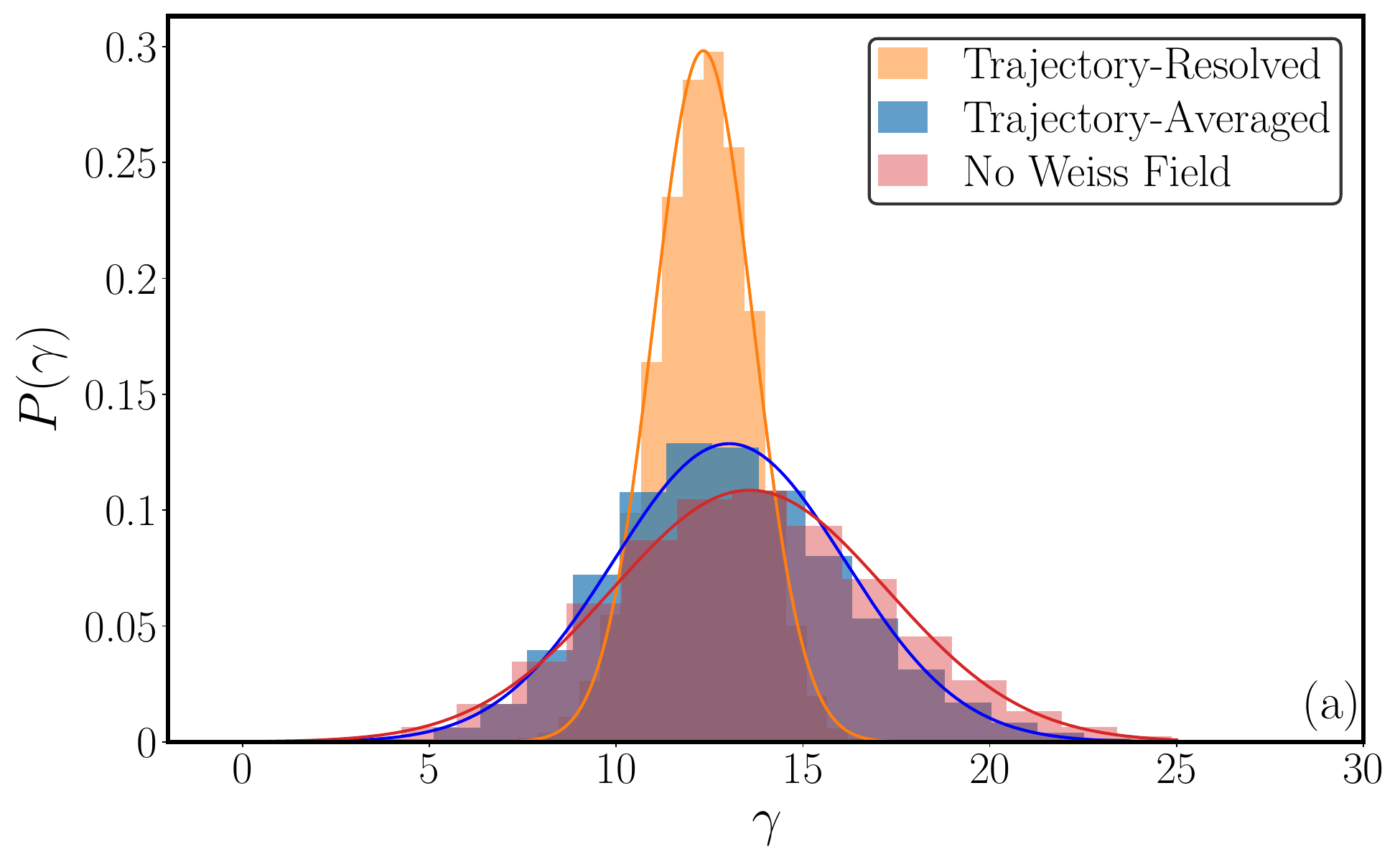}}

\subfloat{\includegraphics[width = 8.8cm]{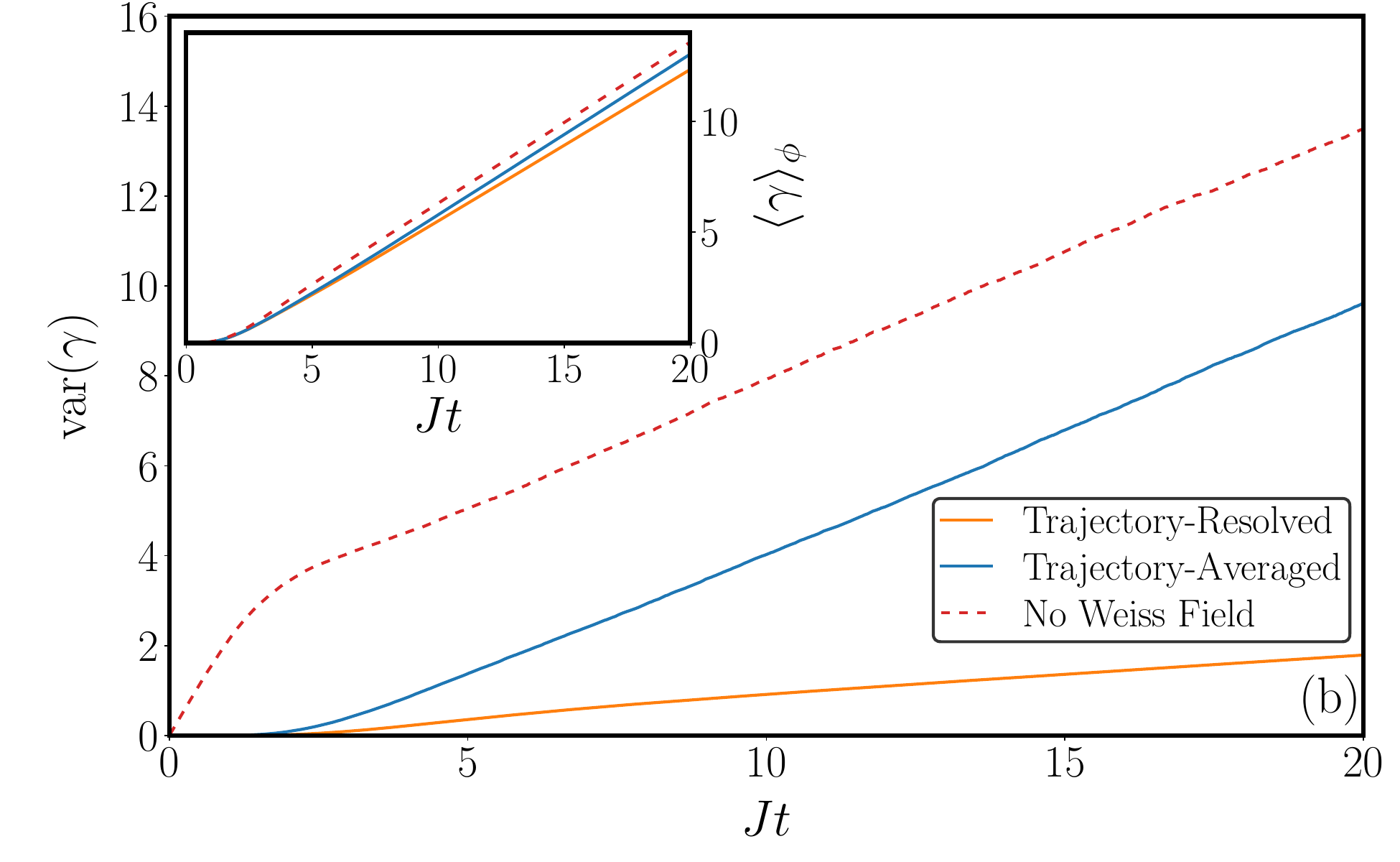}}
\caption{(a) Distribution $P(\gamma)$ at a fixed time $Jt = 20$ following a quench in the one-dimensional quantum Ising model with $N = 9$. We start in the fully-polarized state $\ket{\Downarrow}$ and quench to $\Gamma/J = 0.3$, taking $\mathcal{N} = 5 \times 10^4$ stochastic samples. The use of a trajectory-resolved Weiss field (orange) results in a narrower distribution than the trajectory-averaged case (blue) and the case without a Weiss field (red).  The distributions are approximately normal, as indicated by the solid line fits. (b) Time-evolution of the variance $\text{var}(\gamma)$. The variance grows linearly following an initial transient. The growth rate for the trajectory-resolved case is lower than the other two cases. 
(c) Time-evolution of the mean of the distribution $\langle \gamma \rangle_{\phi}$. The linear growth rate is similar for all three cases.} 
\label{fig:dist2_1D}
\end{figure}

In Fig. \ref{fig:dist2_1D} we consider the distribution of $\gamma = \ln W$, where $W$ is the norm of the stochastic state discussed in Section \ref{sec:stochasticnorm}  of the main text, for the 1D quantum Ising model with $N = 9$ spins. As found in Fig. \ref{fig:dist2} of the main text for the 2D case, the use of a trajectory-resolved Weiss field results in a narrower distribution of $\gamma$ with a slower growth of the variance. This results in improvements in the sampling efficiency for quantum dynamics, even in 1D.

\clearpage

\end{document}